\documentclass[10pt,a4paper,twocolumn,preprintnumbers,citeautoscript,
               aps,prd,longbibliography,superscriptaddress,
               floatfix,nofootinbib]{revtex4-2}

\UseRawInputEncoding
\usepackage{comment}
\usepackage[normalem]{ulem}

\usepackage{amsmath,amsfonts,amssymb}
\usepackage{mathtools}
\usepackage{mathrsfs}
\usepackage{bbm}
\usepackage{braket}
\usepackage{slashed}
\usepackage{siunitx}

\usepackage{graphicx}
\usepackage{adjustbox}
\usepackage{cancel}

\usepackage{booktabs}
\usepackage{makecell}

\usepackage{tikz}
\usetikzlibrary{positioning,shapes,arrows,
                decorations.pathmorphing,
                decorations.markings}

\usepackage[dvipsnames,svgnames]{xcolor}

\usepackage[utf8]{inputenc}
\usepackage[T1]{fontenc}
\usepackage{lmodern}

\usepackage{csquotes}

\usepackage{standalone}

\usepackage[bookmarks=false, breaklinks]{hyperref}

\definecolor{nicered}{rgb}{0.5,0.0,0.0}
\definecolor{darkblue}{rgb}{0.0,0.1,0.9}
\definecolor{lightblue}{rgb}{0.0,0.1,0.6}
\definecolor{applegreen}{rgb}{0.55, 0.71, 0.0}
\definecolor{darkgreen}{rgb}{0.0, 0.2, 0.13}

\hypersetup{
    colorlinks=true,
    citecolor=darkgreen,
    linkcolor=nicered,
    urlcolor=lightblue
}

\begin{document}

\title{Constraints on a Light Leptophilic Scalar from Dark-Sector Couplings}

\author{Marco Graziani}
\email{marco.graziani5@unibo.it}
\affiliation{Physics Department and INFN Sezione di Roma,
Universit\`a di Roma La Sapienza, P.le A. Moro 2, I-00185 Roma, Italy}

\author{Giacomo Landini}
\email{giacomo.landini@lnf.infn.it}
\affiliation{Istituto Nazionale di Fisica Nucleare, Laboratori Nazionali di Frascati, C.P. 13, 00044 Frascati, Italy}

\author{Federico Mescia}
\email{federico.mescia@lnf.infn.it}
\affiliation{Istituto Nazionale di Fisica Nucleare, Laboratori Nazionali di Frascati, C.P. 13, 00044 Frascati, Italy}

\author{Claudio Toni}
\email{claudio.toni@lapth.cnrs.fr}
\affiliation{LAPTh, Universit\'e Savoie Mont-Blanc et CNRS, 74941 Annecy, France}

\author{Ludovico Vittorio}
\email{ludovico.vittorio@uniroma1.it}
\affiliation{Physics Department and INFN Sezione di Roma,
Universit\`a di Roma La Sapienza, P.le A. Moro 2, I-00185 Roma, Italy}

\begin{abstract}

We study a minimal framework where a Majorana fermion dark matter particle interacts with a light scalar mediator coupled mainly to electrons. 
We examine both  freeze-out and freeze-in production to determine the regions of parameter space that yield the correct relic abundance with particular emphasis on a detailed comparison with results in the recent literature. The analysis includes cosmological and astrophysical constraints as well as laboratory bounds from electron-recoil experiments, fixed target searches, and precision measurements. 
The combined results identify a narrow but viable parameter region, favoring sub-GeV dark matter, and define clear targets for future experimental tests. This highlights the strong complementarity between direct‑detection experiments and collider searches.
We additionally investigate the mediator-mass region around 17 MeV, motivated by the  hints reported by the ATOMKI experiment and the PADME collaboration, including couplings between the mediator and light quarks. Direct searches already constrain a large region of the parameter space, even when dark matter is produced via freeze-in, pointing again to sub-GeV dark matter. 

\end{abstract}

\twocolumngrid
\maketitle

\section{Introduction}\label{sec:intro}

Understanding the particle nature of Dark Matter (DM) remains one of the most compelling open questions in modern physics. Among the theoretically motivated possibilities, models in which DM interacts preferentially with leptons have attracted growing interest. Such \emph{leptophilic} scenarios naturally evade the strong bounds associated with hadronic processes while offering clean signatures in low-energy and fixed-target experiments~\cite{Chen:2018vkr,Kopp:2009et,Horigome:2021qof,Jia:2021mwk,Freitas:2014jla,Gu:2017gle,Asai:2022uix,Rella:2022len,BaBar:2020jma,Zhu:2021vlz,Chen:2025ewx, Fox:2008kb,Cao:2009yy,Ibarra:2009bm,Bai:2014osa, Okawa:2020jea, Gninenko:2014pea, YaserAyazi:2019psw,Armando:2023zwz, Athron:2017drj,Tang:2025vqf,Abdughani:2021oit, Koechler:2025ryv, Cesarotti:2024rbh, Kolay:2025jip}.

Within this broad class, an especially relevant limit is that of \emph{electrophilic} interactions, where the mediator couples primarily to electrons. This choice is further motivated by several considerations. Electrons are ubiquitous in both laboratory detectors and astrophysical environments, leading to enhanced production rates and detection sensitivity, particularly for light mediators. In contrast, the heavier and less abundant muons and tau leptons suppress experimental accessibility at low energies, while scenarios involving muon couplings are increasingly constrained by precision observables and tau couplings remain comparatively weakly probed. Electron interactions are theoretically clean, being largely free from hadronic uncertainties, and can be tested across a wide range of complementary settings. Moreover, the growing sensitivity of low-threshold detectors to electron recoils provides powerful probes of such scenarios. Altogether, restricting the interaction to electrons yields a minimal, predictive, and experimentally accessible framework, enhancing the overall testability of the model.

While different mass ranges can be explored, it is particularly well motivated  to consider scenarios in which DM couples to a light mediator with a mass in the $10-100$ MeV interval.
On the one hand, such light mediators attracted interest as they can give rise to sizable elastic DM self-interaction cross sections which could reconcile the cuspy central densities predicted by $N$-body simulations of collisionless cold DM~\cite{Dubinski:1991bm,Navarro:1995iw,Navarro:1996gj} with a range of astrophysical observations that instead favor shallower, core-like profiles in small astrophysical halos~\cite{Dave:2000ar,Vogelsberger:2012ku,Rocha:2012jg,Peter:2012jh,Elbert:2014bma,Fry:2015rta}. Models featuring light mediators are subject to a wide range of experimental and cosmological constraints, which severely restrict the viable parameter space.
In the electrophilic scenario some of these constraints, such as those arising from dark matter direct detection are a priori potentially weakened.

On the other hand, the same range of mediator masses largely overlaps with the target of the broad experimental program dedicated to searching for light particles coupled to electrons, providing an additional strong motivation for considering light electrophilic mediators.  These efforts include, among others, BDX and E12-21-003 at JLab~\cite{Dutta:2023ifr,Celentano:2022ehx,Battaglieri:2022dcy}, LDMX at SLAC and Fermilab~\cite{LDMX:2025bog}, PADME at LNF~\cite{PADME:2022fuc}, as well as NA64~\cite{NA64:2024klw} and NA62~\cite{Dobrich:2018ezn} at CERN. 
While not exhaustive, this list of experiments illustrates the growing interest in probing light, weakly coupled states.
It is therefore important to identify a theoretically motivated and well-defined parameter space to guide and interpret these searches.

Within this broader context, the PADME collaboration explores the production of new light particles coupled to electrons from resonant electron-positron annihilations. Intriguingly, recent data suggest a possible excess in $e^+e^-$ annihilation around a center-of-mass energy of $\sqrt{s}\sim 17$ MeV~\cite{PADME:2025dla}. Although the global significance is only at the $\sim2\sigma$ level, the observation is intriguing because it is consistent, within uncertainties, with the earlier ATOMKI tensions observed in nuclear transitions~\cite{Krasznahorkay:2015iga,Krasznahorkay:2018snd,Krasznahorkay:2019lyl,Krasznahorkay:2021joi,Krasznahorkay:2022pxs,Krasznahorkay:2023sax,Krasznahorky:2024adr}, which, however, requires the presence of hadronic couplings.
These hints reported in nuclear transitions of light nuclei by the ATOMKI collaboration seem to suggest the existence of a new particle with mass around $17$ MeV, commonly referred to as $X_{17}$.
If such a state exists and couples to electrons, it naturally provides a minimal and well-motivated portal between the SM and a light dark sector. In this work, we explore this possibility further by extending our electrophilic analyses to incorporate couplings to first-generation quarks. We highlight the impact of such hadronic couplings  from the perspective of  DM phenomenology. 

The paper is organized as follows. In Sec.~\ref{sec:model}, we introduce the theoretical framework for a scalar boson coupled to electrons that mediates interactions between the SM and the dark sector. 
In Sec.~\ref{sec:BoltzmannOUT}, we show how this setup can reproduce the observed DM abundance via thermal freeze-out. Phenomenological constraints relevant to the freeze-out scenario are discussed in Sec.~\ref{sec:pheno}, and our main results for this case are summarized in Sec.~\ref{sec:results}.
We then examine in Sec.~\ref{sec:X17OUT} the impact of hadronic couplings in the case of the $X_{17}$ boson showing that thermal dark matter production mediated by $X_{17}$ is ruled out by existing data. Finally, we focus on non-thermal DM production via the freeze-in mechanism for both an electrophilic mediator and the $X_{17}$ boson in Secs.~\ref{sec:electroFI} and \ref{sec:X17FI}, respectively.
Finally, Sec.~\ref{sec:conclusions} summarizes our main findings and outlines prospects for future experimental tests of this scenario.

\section{DM WITH LEPTOPHILIC MEDIATOR}
\label{sec:model}

In this work, we investigate a minimal scenario in which fermionic dark matter interacts via a CP-even scalar mediator. The spin and parity of the mediator are chosen such that the dark matter annihilation cross section is velocity suppressed (\(p\)-wave). This feature allows the model to evade stringent constraints from indirect detection searches and observations of the Cosmic Microwave Background (CMB).\footnote{For scalar dark matter, \(p\)-wave suppression can instead be achieved with a vector mediator.}

The low-energy interactions of a Majorana\footnote{A Dirac fermion would lead to qualitatively similar results.} DM fermion $\psi$ and a CP-even leptophilic scalar mediator $X$ are described by the Lagrangian
\begin{equation}\label{eq:LAG}
    \mathcal{L}_{\rm int}=\sum_\ell g_{\mathcal{\ell}}\bar{\ell}\ell X+\frac{g_\psi}{2}{\psi}\psi X \,,
\end{equation}
with $\ell=e,\mu,\tau$ denoting the SM leptons. Since the Lagrangian is not invariant under the electroweak gauge group, a proper UV completion is needed to generate the interactions with the SM leptons.
For instance, minimal UV completions in which the Standard Model (SM) field content is augmented by a heavy vector-like copy of the right-handed electron field $e_R$ can lead to Eq.~\eqref{eq:LAG}.
We provide an example of explicit UV completion to Appendix~\ref{app:UV}, whereas in the following we focus on the phenomenological implications of Eq.~\eqref{eq:LAG} in a model-independent way.

In a generic model, the couplings $g_{\ell}$ to the different lepton families are unrelated. However, specific scenarios motivated by UV physics are typically considered in the literature. These include universal couplings, $g_e = g_\mu = g_\tau$, mass-proportional couplings,
$g_\ell = (m_\ell/m_e)\, g_e$, or couplings to only one or two specific generations, such as electrophilic or muon-philic scenarios~\cite{Chen:2018vkr,Kopp:2009et,Horigome:2021qof,Jia:2021mwk,Freitas:2014jla,Gu:2017gle,Asai:2022uix,Rella:2022len,BaBar:2020jma,Zhu:2021vlz,Chen:2025ewx,Fox:2008kb,Cao:2009yy,Ibarra:2009bm,Bai:2014osa,YaserAyazi:2019psw,Athron:2017drj,Okawa:2020jea,Gninenko:2014pea,Medina:2021ram,Tang:2025vqf,Abdughani:2021oit,Koechler:2025ryv,Cesarotti:2024rbh}.

In the following, we  discuss an electrophilic scalar mediator driven by the coupling $g_e$.
Including couplings to muons and taus does not quantitatively modify the DM phenomenology of our setup. However, the corresponding collider and low-energy constraints would require a more dedicated and involved analysis~\footnote{See, e.g., Ref.~\cite{Cesarotti:2024rbh} under the 
assumptions  \( m_X > m_\psi \) and $g_\ell = (m_\ell/m_e)\, g_e$}, which we leave for future work. 

To explore the DM phenomenology  of the model in Eq.~\eqref{eq:LAG}, the mass hierarchy between the mediator $X$ and the DM particle $\psi$ is crucial.

The case of heavy mediators, \( m_X > m_\psi \), has been discussed in Ref.~\cite{Cesarotti:2024rbh}. For sufficiently large values of the coupling $g_\ell$, the DM particle is kept in thermal equilibrium with the SM bath in the early Universe. When the bath temperature drops below the DM mass, the latter undergoes thermal freeze-out driven by annihilations to SM leptons mediated by $X$, namely $\psi\psi\to X\to \bar{\ell}\ell$.
Requiring the DM relic abundance to match the observed value, $\Omega_{\rm DM} h^2 = 0.1200 \pm 0.0012$~\cite{Planck:2018vyg}, determines the value of the coupling $g_\ell$.
Several conclusions can be drawn. First, the purely electrophilic scenario is ruled out. Indeed, the values of the coupling $g_e$ needed to reproduce the DM relic abundance are excluded by direct detection (DD) and collider experiments for DM masses $m_\psi\gtrsim 5$ MeV. 
Lighter DM candidates are excluded by Big Bang Nucleosynthesis (BBN) constraints, since they would undergo freeze-out at late times, resulting in a large thermal abundance at temperatures $T \sim \mathcal{O}({\rm MeV})$, which would affect the primordial abundances of light elements. 

In a more general setup, in which $X$ couples to all three lepton families, the model remains excluded for DM masses below the muon threshold, $m_\psi < m_\mu$, for the same reasons discussed above. Let's stress that the current DD constraints on the DM-electron scattering cross section are stronger than those used in Ref.~\cite{Cesarotti:2024rbh}.
Retaining the assumption $m_X > m_\psi$, and considering scenarios in which the coupling to muons is absent or suppressed in light of recent results on $(g_\mu - 2)$~\footnote{
{
Unlike in the past, the current determination of the anomalous magnetic moment of the muon~\cite{Aliberti:2025beg},
$\Delta a_\mu \equiv a_\mu^{\rm exp} - a_\mu^{\rm SM} = (38 \pm 63)\times 10^{-11}$, where \(a_\mu \equiv (g_\mu - 2)/2\), 
shows no evidence of new physics that would motivate a coupling to the muon.
}}, the observed DM relic abundance can be reproduced only above the $\tau$ threshold, $m_\psi > m_\tau$, provided that the coupling to the $\tau$ lepton is sufficiently large.

{This conclusion changes drastically once the inverse mass hierarchy $m_\psi > m_X$ is assumed. 
In the following, we focus on this alternative scenario in the case of an electrophilic scalar mediator.
}

In this setup,  we consider thermal freeze-out in the regime $g_\psi \gg g_e$, so that DM particles predominantly annihilate into the light mediator through the secluded channel $\psi\psi \to XX$. This mechanism efficiently reduces the DM abundance and opens up a sizable and previously inaccessible region of parameter space. Although secluded annihilations may appear less predictive, they nevertheless give rise to rich and testable phenomenological implications.

First, the relatively large values of $g_\psi$ required by thermal freeze-out, combined with the lightness of the mediator, can lead to observable astrophysical signatures, in particular through sizable DM self-interactions. On the other hand, while the electron coupling $g_e$ does not directly control the freeze-out dynamics, it plays a crucial role in experimental probes. In particular, stringent constraints can be derived from DD experiments sensitive to elastic DM-electron scattering. These bounds can be compared with complementary constraints from collider and low-energy experiments, which already impose non-trivial upper limits on $g_e$.

In this respect, LEP searches, including both decay-mode-independent scalar searches and analyses based on invisible final states, significantly constrain scenarios involving light scalar electrophilic mediators~\cite{LEP:InvisibleHiggs2001,LEP:ScalarS0}. In addition, the parameter space of electrophilic scalar models has been probed by the Belle~\cite{Belle:2022gbl} and BaBar~\cite{BaBar:2020jma} collaborations, as well as by beam-dump experiments~\cite{Bjorken:1988as,Davier:1989wz}, through searches for processes of the type $e^+e^- \to e^+e^-(X\to e^+e^-)$. 

Furthermore, a dedicated analysis of Belle data~\cite{Belle:2022gbl} assuming a universally coupled leptophilic scalar $X$ was performed in Ref.~\cite{Cogollo:2024fmq}. This study yields strong constraints on $g_e$  from $e^+e^-\to e^+e^-(X\to e^+e^-)$, which can be straightforwardly applied to the electrophilic scenario considered here.

\section{Thermal freeze-out}
\label{sec:BoltzmannOUT}

When the coupling of the light mediator $X$ to the Majorana fermion $\psi$ is sizable, elastic scatterings and annihilations keep the two particles in kinetic and chemical equilibrium in the early Universe. Moreover, both of them thermalize with the SM thermal bath as long as the rates of decays and inverse decays of $X$ to SM leptons are faster than the Hubble expansion rate $H=\sqrt{4\pi^3 g_\rho/45}T^2/M_{\rm Pl}$, being $T$ the temperature of the SM thermal bath and $g_\rho$ the number of relativistic degrees of freedom ($g_\rho\sim 10.75$ within the SM at $T\sim 10$ MeV). More precisely, the mediator decays with rate $\Gamma_X=\sum_{\ell}\Gamma_{X\to \ell \ell}$, where
\begin{equation}
    \Gamma_{X\to\ell \ell}=\frac{g_\ell^2 m_X}{8\pi}\left(1-\frac{4m_\ell^2}{m_X^2}\right)^{3/2}\,,
\end{equation}
so that thermal equilibrium with the SM bath is maintained if
\begin{equation}
    \frac{\Gamma_X\, m_X}{4H T}e^{-2m_X/T}\bigg|_{T=T_\text{fo}}\gtrsim 1\,,
\end{equation}
where $T_\text{fo}\simeq m_\psi/25$ indicates the typical temperature for DM freeze-out, see below. For a mediator of mass $m_X\sim10$ MeV, this corresponds to a lower bound on the lepton coupling $g_\ell\gtrsim 10^{-\{9,8,7\}}$ for DM masses $m_\psi\sim \{0.1,1,10\}\text{ GeV}$.~\footnote{Notice values of $g_e$ below those probed by the experiment E137~\cite{Bjorken:1988as} do not thermalize and are therefore not relevant, see Fig.~\ref{fig:DMparamspace}.} On the other hand, for a milder hierarchy, $m_X\lesssim m_\psi$, thermalization is much less efficient due to the exponential suppression and larger couplings are needed. We found that, as long as $m_\psi\gtrsim 2m_X$, thermalization is achieved for values of the coupling $g_e\gtrsim 10^{-6}$ in all the mediator mass range we are interested in.

The DM particles annihilate to $XX$ and ${\ell^+}\ell^-$. In the non-relativistic limit the relevant cross sections are  
\begin{equation}\label{eq:XX}
        \sigma v_{\psi\psi\to XX}=\dfrac{g_{\psi}^{4}\,\sqrt{1-\dfrac{m_{X}^{2}}{m_{\psi}^{2}}}}{24\pi\,m_{\psi}^{2}\left(2-\dfrac{m_{X}^{2}}{m_{\psi}^{2}}\right)^{4}}
        \left(\dfrac{2m_{X}^{4}}{m_{\psi}^{4}}-\dfrac{8m_{X}^{2}}{m_{\psi}^{2}}+9\right) \,v^2 
\end{equation}
and 
\begin{equation}\label{eq:ll}
    \sigma v_{\psi\psi\to \ell^+\ell^-}=\frac{g_{\psi}^{2}\,g_\ell^{2}m_\psi^2\left(1-m_{\ell}^{2}/m_\psi^2\right)^{3/2}}
{8\pi\left[\left(m_{X}^{2}-4m_{\psi}^{2}\right)^{2}+\Gamma_{X}^{2}m_{X}^{2}\right]}\,v^2\,,
\end{equation}
where $\Gamma_X$ is the finite width of the $X$-mediator appearing in the $s$-channel propagator. Notice that both cross sections are $p$-wave suppressed, as expected for the annihilation of Majorana and Dirac  fermions mediated by a scalar. The velocity suppression is crucial, since it allows us to evade constraints from CMB and indirect detection searches, see below. We assume $g_\psi\gg g_\ell$, so that the $\ell^+\ell^-$ channel is subdominant and the relic density is controlled by the coupling $g_\psi$.

As long as DM annihilations are faster than the Hubble expansion, the DM particle follows its equilibrium distribution. This holds until the Universe cools down to the freeze-out temperature $T\sim m_\psi/25$, when the annihilation rate drops below Hubble and the DM abundance freezes to its present value. More precisely, the DM relic abundance is obtained solving the Boltzmann equation 
\begin{equation} 
\label{eqBoltz}
\dot{n}_\psi+3Hn_\psi=-\langle{\sigma v}\rangle_{\psi\psi}(n_\psi^2-n_{\psi,\rm eq}^2)\,,
 \end{equation}
where $n_\psi$ is the DM number density and $n_{\psi,\rm eq}=(m_\psi^2 T/\pi^2)K_2(m_\psi/T)$ its equilibrium value. Here $\sigma v$ is the sum of the contributions from Eq.~\eqref{eq:XX} and Eq.~\eqref{eq:ll}, while $\langle{\rangle}$ denotes the thermal average.
As discussed before, the second contribution is subdominant for $g_\ell\ll g_\psi$ and we neglect it in the following.  Parameterizing the annihilation cross section as $\sigma v=\sigma _1 v^2$ and using $\langle{v^2\rangle}=6T/m_\psi$, the solution of the Boltzmann equation can be analytically approximated as 
\begin{equation}
    \frac{\Omega_{\rm DM}h^2}{0.12}\simeq \left(\frac{x_{\rm fo}}{25}\right)^2\frac{1.8\times 10^{-25}\text{cm}^3/\text{sec}}{\sigma_1}\,,
\end{equation}
where $x_{\rm fo}=m_\psi/T_\text{fo}$.
Then, reproducing the observed cosmological relic density fixes the value of the coupling $g_\psi$ as a function of the DM mass as $g_\psi\simeq 10^{-3}\,\sqrt{m_\psi/\text{MeV}}$. We checked this result by solving numerically Eq.~\eqref{eqBoltz} for different values of $m_\psi$ and $g_\psi$. We underline the characteristic dependence on the DM mass, $g_\psi^2\propto m_\psi$. 

If the mediator is sufficiently light, the annihilation cross sections in Eqs.~\eqref{eq:XX} and~\eqref{eq:ll} are enhanced by the so-called Sommerfeld factor, which takes into account the deflection of the wave-function of the initial state from a plane wave (see for instance~\cite{Mitridate:2017izz}). This effect becomes relevant when $\alpha_\psi m_\psi/m_X\gtrsim1$, where $\alpha_\psi=g_\psi^2/4\pi$.
The maximum enhancement typically arises at low Dark Matter velocities, so that such effect is particularly significant for indirect detection searches, even if the cross section is $p$-wave suppressed. 
In agreement with the results of Ref.~\cite{Tulin:2013teo}, we find that the Sommerfeld factor becomes important for DM masses $m_\psi\gtrsim300$ GeV, modifying the relation between $g_\psi$ and $m_\psi$. 

Additionally, the exchange of a light mediator can lead to the formation of DM bound states. Bound state formation (BSF) arises mainly via radiative capture, when two DM particles form a bound state and release the corresponding binding energy by emitting a light mediator. It turns out that, in the presence of a scalar mediator, the contribution at $\mathcal{O}(\alpha_\psi^2)$ vanishes and the leading cross section for BSF arises only at $\mathcal{O}(\alpha_\psi^4)$. As a consequence, BSF becomes relevant only for values of $\alpha_\psi$ typically much larger than those required to reproduce the observed DM relic abundance (at least for DM masses below a few TeV), and it can therefore be safely neglected in the discussion of freeze-out, which is largely model-independent.  For more details on BSF, see Refs.~\cite{Wise:2014jva,Petraki:2015hla,Petraki:2016cnz,Oncala:2018bvl}. On the other hand, BSF can be very relevant at the time of recombination and may provide strong constraints from CMB observations for sufficiently heavy DM, as discussed below.

In the remainder of this Section, we revert to the case $\ell = e$; the discussion of freeze-out presented above is however generic and does not rely on this specific choice.

\subsection{Phenomenological constraints}
\label{sec:pheno}

\subsubsection{CMB and Indirect detection}

Dark Matter annihilations $\psi\psi\to XX$ are followed by the instantaneous decays $X\to e^+e^-$ and are  potentially constrained by CMB measurements and indirect detection searches. However, due to their $p$-wave nature, the corresponding cross sections are suppressed by the low DM velocity both at the time of recombination and at present time, so that the constraints are safely evaded in a large portion of the parameter space. The same argument applies to direct $\psi\psi\to e^+e^-$ annihilations, which are anyway sub-leading.

It is important to point out that the statement above loses validity in the region of the parameter space where Sommerfeld enhancement and bound state formation become relevant, in particular for heavy DM masses and low DM velocities, such as the ones at the time of recombination, $v\sim 10^{-8}$. Indeed, while being sub-dominant at the time of DM freeze-out, the cross sections for BSF are not velocity suppressed. BSF via radiative capture is accompanied by the emission of a light mediator, which subsequently decays to electrons, leading to an unsuppressed signals. Following the analysis of Ref.~\cite{An:2016kie}, the entire region of the parameter space where BSF formation is kinematically allowed, $\alpha_\psi^2m_\psi/4m_X>1$, is either ruled out or disfavored by CMB data. Within the assumption of thermal production, this corresponds to DM masses $m_\psi\gtrsim 300$ GeV. 

\subsubsection{Bullet Cluster and SIDM}

Scenarios with a DM elastic self-interaction cross section per unit mass of $\mathcal{O}(\text{cm}^2/\text{gr})$ 
are known in the literature as Self-Interacting Dark Matter (SIDM) models~\cite{Spergel:1999mh}. These attracted attention due to their ability to reduce the central densities of DM halos. Indeed, such large cross sections could alleviate some tension between the predictions from simulations of collisionless cold DM~\cite{Dubinski:1991bm,Navarro:1995iw,Navarro:1996gj} and certain astrophysical observations from dwarf galaxies and other small-scale structures~\cite{Dave:2000ar,Vogelsberger:2012ku,Rocha:2012jg,Peter:2012jh,Elbert:2014bma,Fry:2015rta}.
On the other hand, observational data from galaxy clusters, such as the Bullet Cluster,
impose stringent upper bounds on the DM elastic cross section of about $\sigma_{\rm el}/m_{\psi}\lesssim 0.5\text{ cm}^2/\text{gr}$~\cite{Harvey:2015hha,Bondarenko:2017rfu,Harvey:2018uwf,Sagunski:2020spe,DES:2023bzs,ODonnell:2025pkw}.

Models with light mediators can realize the SIDM scenario (see for instance~\cite{Tulin:2013teo,Tulin:2012wi,Loeb:2010gj,Schutz:2014nka,Feng:2009hw,Feng:2016jff,Buckley:2009in,Kahlhoefer:2017umn} as well as~\cite{Tulin:2017ara} and references therein).
Refs.~\cite{Tulin:2013teo,Kahlhoefer:2017umn} identified the regions of the parameter space which provide the correct velocity-dependence and magnitude for DM elastic cross sections, with the additional assumption that the DM relic abundance is obtained with annihilations to the light mediator, $\psi\psi\to XX$, analogously to our scenario, see for example Figure 9 of~\cite{Tulin:2013teo}.\footnote{While that figure refers to Dirac DM, analogous results can be found for Majorana fermions.} For mediator masses in the $10-100$ MeV range, the range of DM masses for viable SIDM lies around $10\text{ GeV}\lesssim m_\psi\lesssim 10\text{ TeV}$. Within this interval, masses larger than $m_\psi\gtrsim 300$ GeV are constrained by CMB data as discussed in the previous section, while the remaining  values are severely constrained by direct detection experiments if the mediator is coupled to quarks at tree level, for instance through an Higgs portal or a dark photon portal (see, e.g.~\cite{Hambye:2019tjt}). The electrophilic scenario of Eq.~\eqref{eq:LAG} could partially alleviate some of these constraints, as coupling to quarks is induced only at two-loops.

More conservatively, abandoning the SIDM paradigm and the requirement of sizable self-interactions in small-scale galaxies and only requiring consistency with the constraints from galaxy clusters, the viable parameter space extends to sub-GeV dark matter.

\subsubsection{Laboratory constraints}

Data from low-energy observables can give important constraints on the parameter space of the CP-even electrophilic scalar mediator $X$, independently of its hypothetical role as a mediator between the visible and the dark sectors. The recent study carried out in Ref.~\cite{DiLuzio:2025ojt} has precisely investigated this kind of bounds and, in what follows, we will briefly review its findings. 

A first constraint on the mediator mass $m_X$ and on its coupling with electrons $g_e$ can be obtained through the study of rare pion decays, namely by searching for the decay channel $\pi^+ \to e^+ \nu_e X$. This is exactly what has been done by the SINDRUM experiment~\cite{SINDRUM:1986klz,SINDRUM:1989qan}, which looked for $e^+e^-$ resonances in $\pi^+ \to e^+ \nu_e X$, followed by the prompt decay of $X$ into an $e^+e^-$ pair. A dedicated search for the same channel will be also developed in the future by the PIONEER Collaboration~\cite{PIONEER:2022yag}.
The constraints labeled as SINDRUM and PIONEER in all the plot contained in the present article have been directly taken from Figure 1 of Ref.~\cite{DiLuzio:2025ojt}.

A second interesting bound on the mediator mass $m_X$ and on its coupling with electrons $g_e$ comes from the anomalous magnetic moment of the electron. Contrarily to the case of the $g-2$ of the muon, which appears to be SM-like, the estimate of $\Delta a_e \equiv a_e^{\rm {exp}} - a_{e}^{\rm {SM}}$ depends on the numerical value of the fine structure constant $\alpha_{\rm em}$. However, recent measurements from Cesium and Rubidium interferometry \cite{Parker_2018, Morel:2020dww}  recently pointed to discrepant values of $\alpha_{\rm em}$ and, thus, of $a_{e}^{\rm {SM}}$. Such a tension then naturally reflects onto the value of $\Delta a_e$. For this motivation, we follow Ref.\,\cite{DiLuzio:2025ojt} in defining two benchmark scenarios, namely $|\Delta a_e| \leq 10^{-12}$ and $|\Delta a_e| \leq 10^{-14}$, in order to infer some bounds from the $g-2$ of the electron. The shift $\Delta a_e^X$ induced by a CP-even scalar particle $X$ reads as~\cite{Jegerlehner:2009ry,DiLuzio:2025ojt}
\begin{equation*}
\Delta a_e^X =  \frac{g_e^2}{4\pi^2} \frac{m_e^2}{m_X^2} L_X,
\end{equation*}
where $L_X \simeq \log[m_X/m_e]-7/12$ for $m_X\gg m_e$.

Further constraints on $g_e$ and $m_X$ can be inferred from experimental searches for electrophilic scalars at BELLE\,\cite{Belle:2022gbl}, BaBar\,\cite{BaBar:2020jma} and beam-dump experiments~\cite{Bjorken:1988as, Davier:1989wz}.

The BELLE collaboration analyzed the process $e^+e^-\to X\bar{\tau}\tau\to \bar{\ell}\ell \bar{\tau}\tau$, which assumes a non-vanishing coupling of $X$ to $\tau$ leptons. 
More recently, Ref.~\cite{Cogollo:2024fmq} re-casted the limits for a scalar particle $X$ with universal couplings to all leptons. For scalar masses between 40 and 200 MeV, the most relevant process is $e^+e^-\to Xe^+e^-\to e^+e^-e^+e^-$ and the limits they obtained can be directly applied to our electrophilic model.
The limit can be further improved by Belle II.
A reliable projection would require a dedicated simulation of the signal
and backgrounds, including the Belle II detector acceptance, selection efficiencies and the
corresponding experimental systematics. A simple estimate of the potential improvement can be obtained
by rescaling the existing limits on cross-sections by Belle corresponding to the increase in
integrated luminosity. This suggests  a potential improvement in the sensitivity to $g_e$ by about one order of magnitude with respect to Ref.~\cite{Cogollo:2024fmq}.

Other bounds on $g_e$ and $m_X$ can also be inferred from electron beam-dump experiments \cite{Bjorken:1988as, Davier:1989wz} (see also the analysis contained in Ref.~\cite{Liu:2016qwd}). In these searches, in fact, an electrophilic scalar is produced in the dump through a bremsstrahlung-like process $e^-N\to e^-N\,X$, followed again by the prompt decay of $X$ into an $e^+e^-$ pair. 

The future polarized positron program at Jefferson Lab (JLab) will provide an additional interesting probe of light mediators coupled to electrons. The corresponding projected bounds have been taken from the recent computations of Ref.~\cite{Pustyntsev:2025iht} and shown in Fig.\,\ref{fig:DMparamspace}. It is worth highlighting that for $m_X \gtrsim 70$ MeV such constraints become competitive with the expected bounds from PIONEER.

One final comment regards the possible improvement in sensitivity to light electrophilic scalars at FCC-ee. Dedicated studies already exist for leptophilic vector mediators~\cite{GonzalezSuarez:2024dsp}. However, these analyses focus on vector mediators and on a mass range significantly heavier than the one considered here, so that their results cannot be straightforwardly translated to our scenario. To the best of our knowledge, no dedicated projection is currently available for a light electrophilic scalar. A reliable estimate of the FCC-ee reach would therefore require a dedicated analysis, which lies beyond the scope of the present work.

\subsubsection{Direct detection--electron scatterings}

Direct detection (DD) experiments are sensitive to scatterings of DM off electrons and provide an upper limit on the corresponding cross section (see for instance~\cite{Essig:2015cda,Essig:2017kqs, Cirelli:2024ssz})\footnote{Recently, Ref.~\cite{Stratman:2026qnh} discusses the constraints from DM scattering off electrons for very light mediators in the keV range for DM produced through the freeze-in mechanism.}. When the DM  particle scatters off the electrons bounded in the atoms of the target material, it can ionize them, provided it possesses enough kinetic energy, i.e. $E\gtrsim 10$ eV.
Experimental limits on the spin-independent (SI) cross section are usually reported in terms of a fiducial cross section $\sigma_{\psi e}$ which, in the limit of heavy mediator, $m_X\gg m_e$, corresponds to the scattering off a free electron. 

The electrophilic interaction of Eq.~\eqref{eq:LAG} predicts
\begin{equation}\label{eq:DMesigma}
    \sigma_{\psi e}\simeq \frac{g_\psi^2g_e^2\mu_{\psi e}^2}{\pi m_X^4}\,,
\end{equation}
which, in the region of the parameter space where the freeze-out dynamics fixes $g_\psi^2\propto m_\psi$, can be re-expressed as  

\begin{equation}\label{eq:sigmanum}
    \sigma_{\psi e}\simeq 2\times 10^{-32}\text{ cm}^2 g_e^2\frac{m_\psi}{100\text{ MeV}}\left(\frac{20\text{ MeV}}{m_X}\right)^4\,.
\end{equation}
The strongest limits are set by the 
XENONnT collaboration, which recently published its more recent results~\cite{XENON:2026qow} for DM masses in the 40 MeV- 10 GeV range.
Since the experimental sensitivity at large DM masses is limited by the DM number density, the limit scales as $\sigma_{\psi e}^{\rm limit}\propto m_\psi$ and can be conservatively extrapolated also to DM masses above 10 GeV.

From Eq.~\eqref{eq:sigmanum}, it is evident that 
DD limits translate into a constraint on the three parameters $\{g_e,m_X,m_\psi\}$.
It is interesting to notice that, for DM masses larger than 1 GeV, both the experimental limit and the theoretical prediction scales linearly with the DM mass, so that the constraint on $g_e$ becomes independent on the precise value of $m_\psi$.

\subsubsection{Direct detection--nuclear scatterings}

Despite the fact that the electrophilic mediator does not couple at tree level with quarks and gluons, such interactions are induced at the loop level, thus providing loop-suppressed DM scatterings off nucleons. These are relevant for DM masses above the GeV, where the experimental limits from nuclear recoils are orders of magnitude stronger than those from electrons.

The coupling to quarks, $g_qX\bar{q}{q}$ is generated via the mixing of $X$ with the Higgs boson, induced by a loop of electrons, roughly estimated as
\begin{equation}
    g_q\sim \frac{y_e y_q g_e}{16\pi^2}\times L_q\,,
\end{equation}
where $y_{q,e}$ are the quark and electron Yukawa couplings, while $L_q$ represents a possible logarithmic enhancement of order $\sim 1-10$. Such induced coupling is too small to give rise to any observable effect.

A second and more dangerous contribution to the DM-nucleus scattering arises via a  two-loop diagram where two photons can be attached to a loop of electrons, connected to the scalar mediator. This two-loop contribution has been extensively discussed in the literature. Ref.~\cite{Kopp:2009et} computed the corresponding DM-nucleus cross section in the approximation of heavy lepton in the loop, which is not completely fulfilled in the electron case. Ref.~\cite{Garani:2021ysl} provides a complete computation, also pointing out the importance of scatterings off multi-nucleons, which become particularly relevant for scatterings off heavy elements such as Xenon. Following these analyses we provide a rough estimate of the upper bound on the coupling $g_e$.

\subsection{Results}
\label{sec:results}

\begin{figure*}[t]
\!\!\!\includegraphics[width=0.5\textwidth]{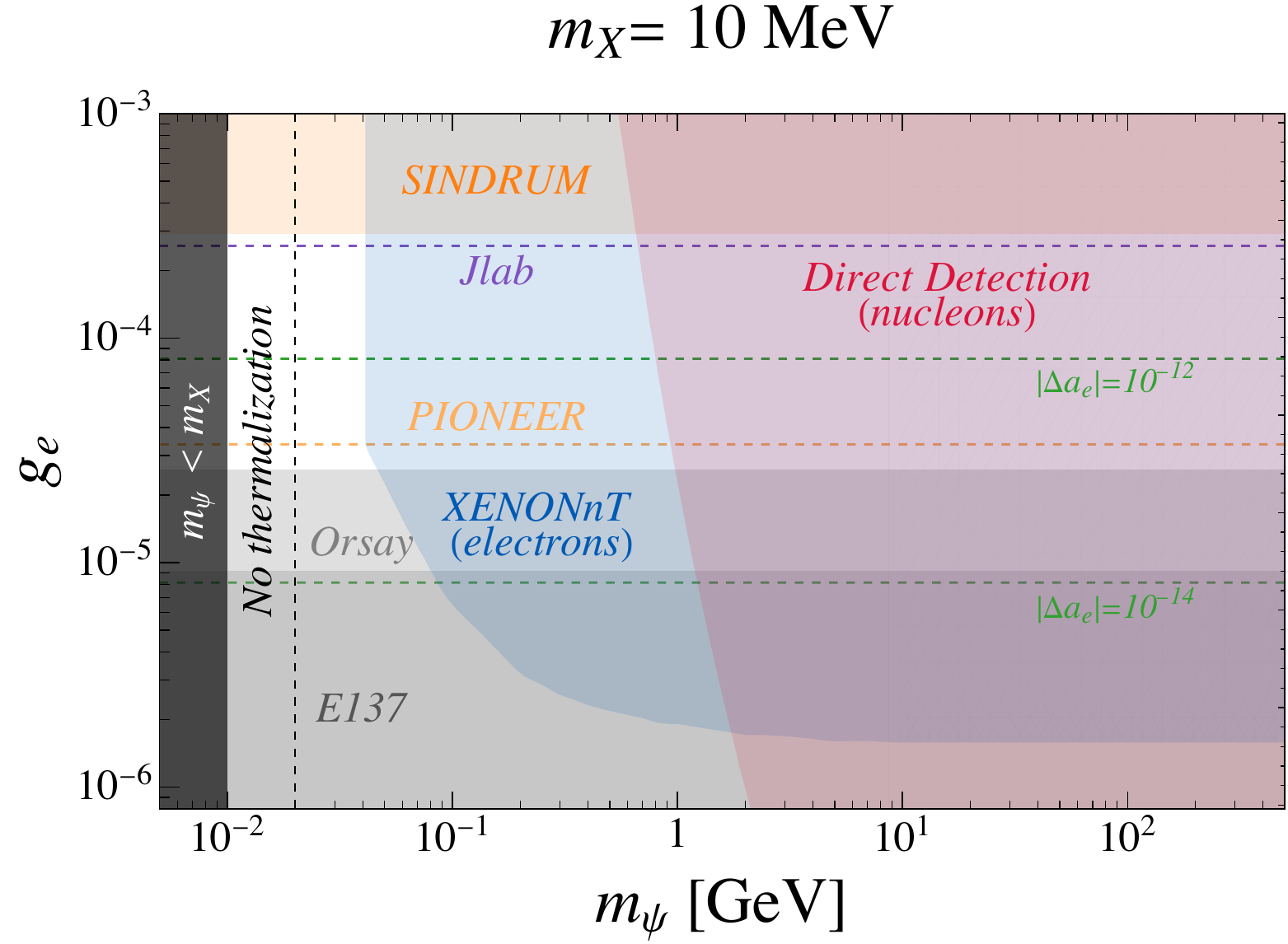}
\includegraphics[width=0.5\textwidth]{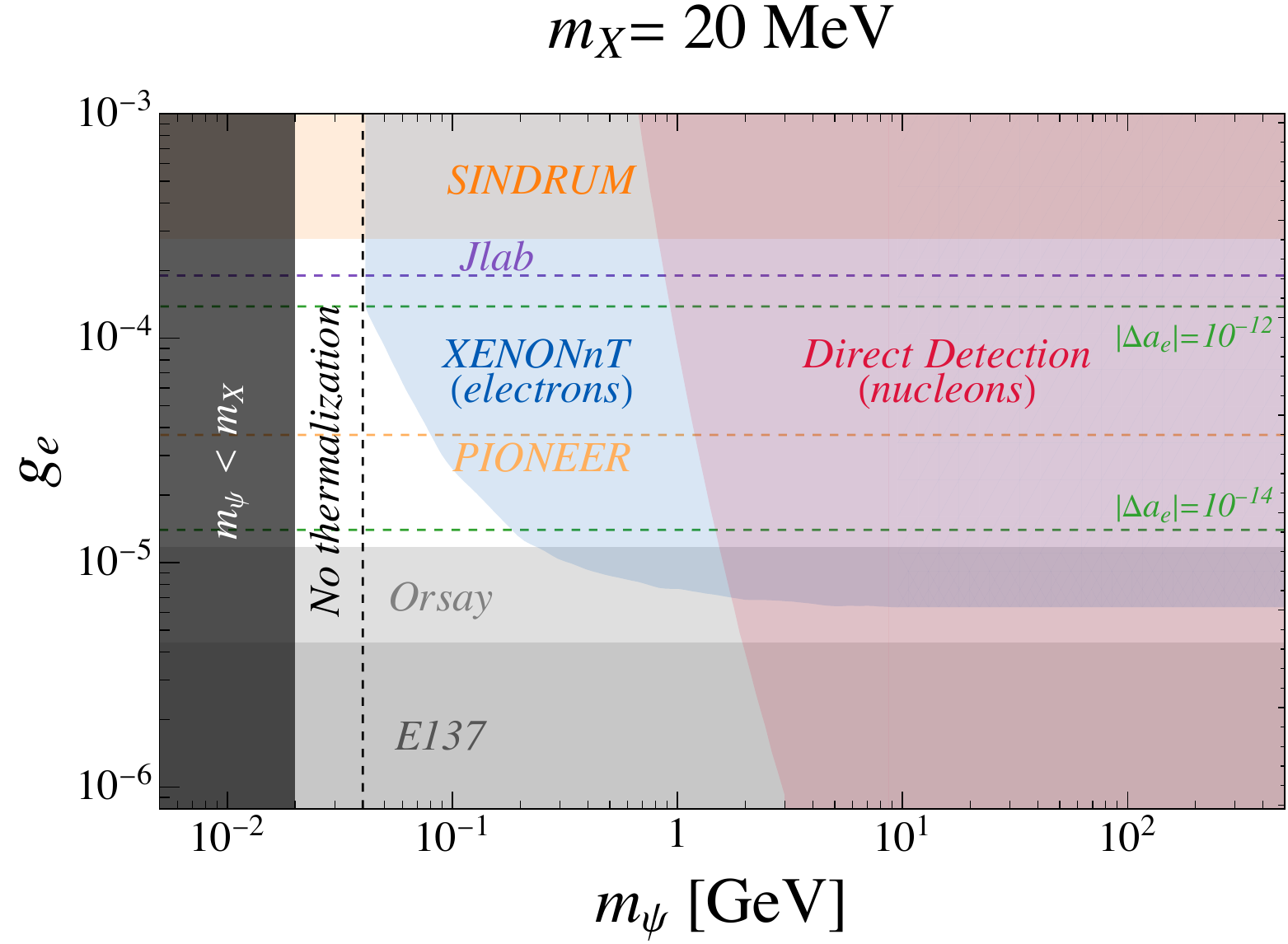}\\[0.4cm]
\!\!\!\includegraphics[width=0.5\textwidth]{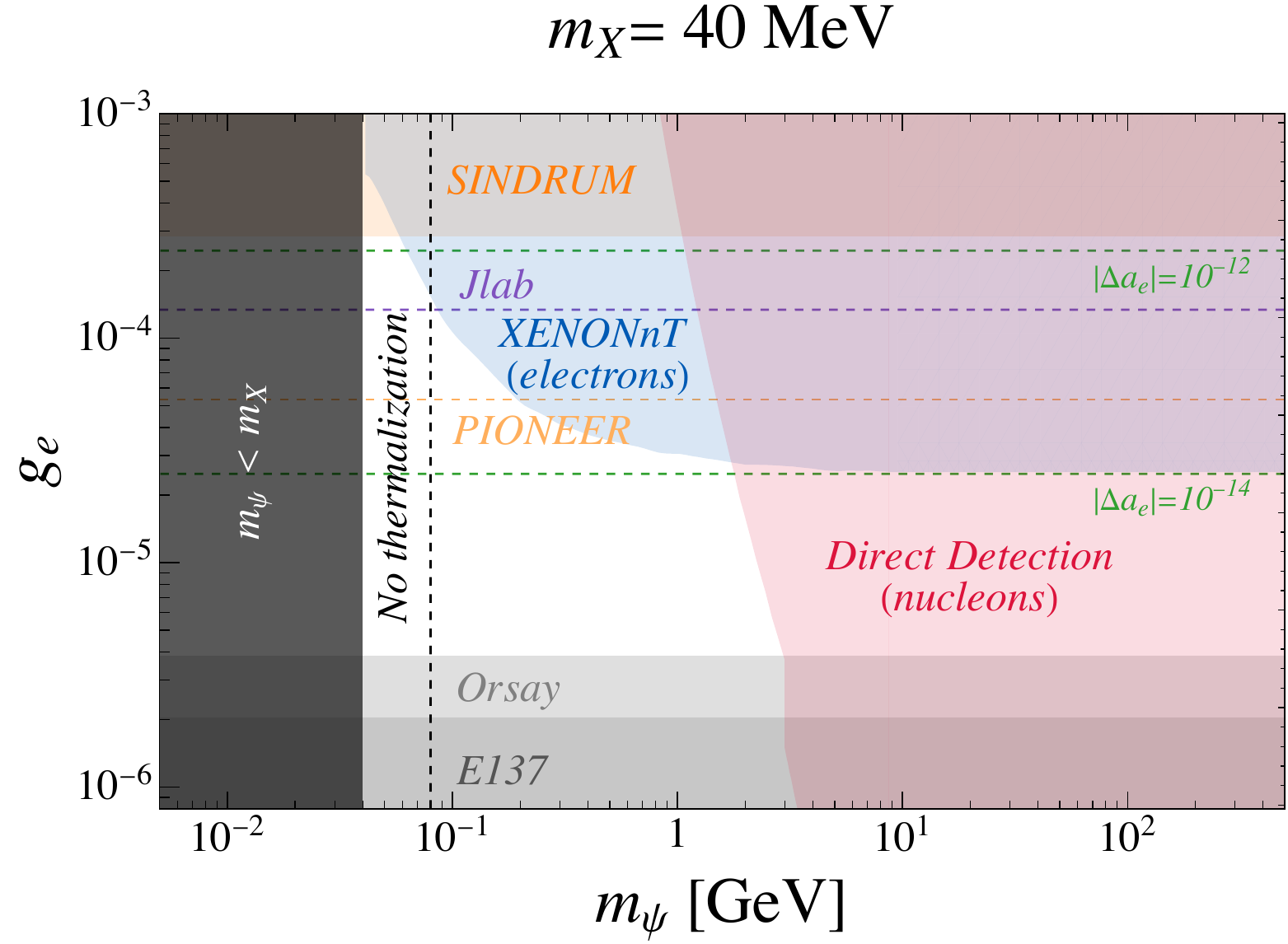}
\includegraphics[width=0.5\textwidth]{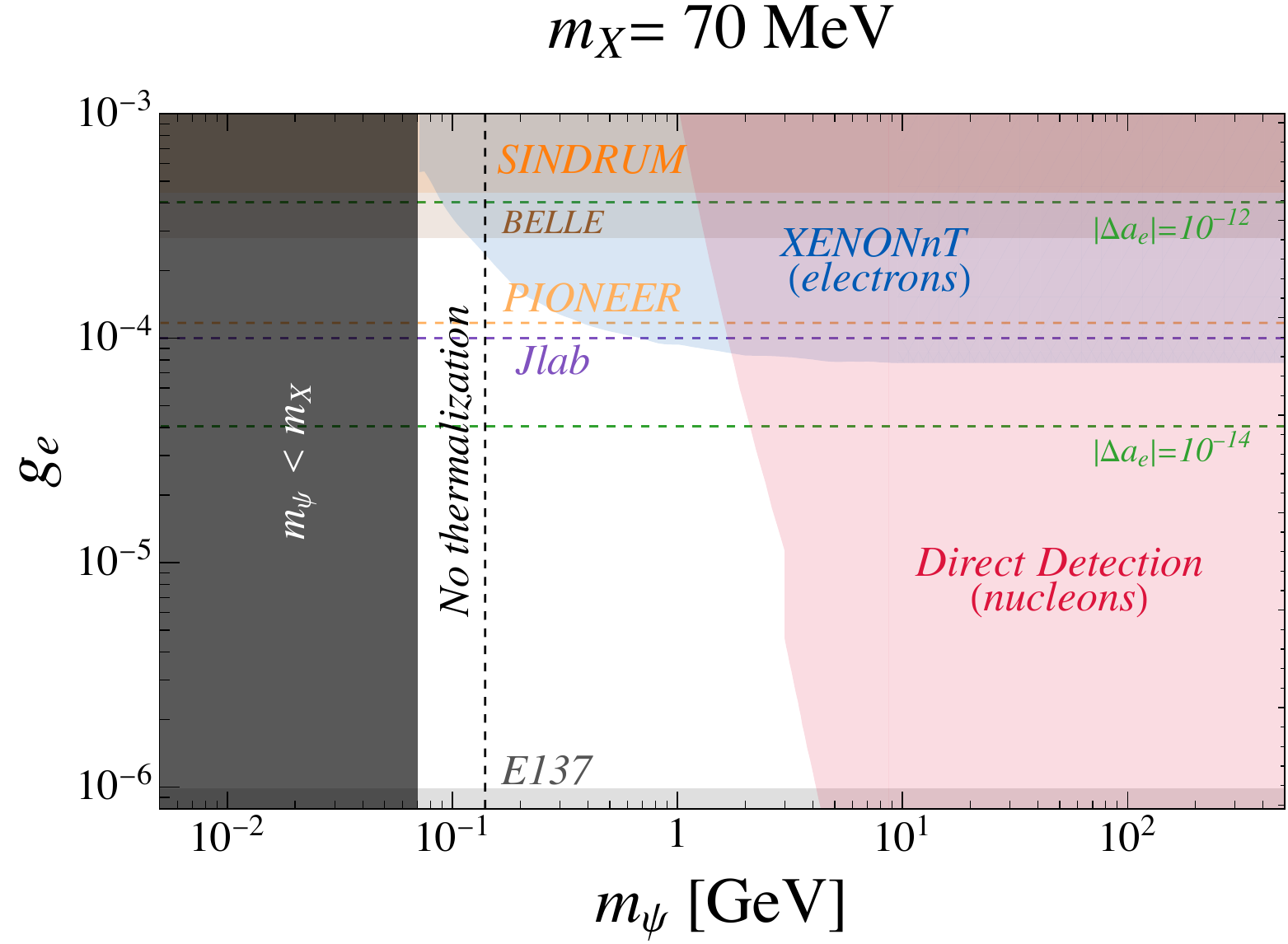}
\caption{Parameter space in the plane $(m_\psi\,,\,g_e)$,
 assuming a CP-even scalar mediator with mass $m_X = 10 ~\mathrm{MeV}$ (top left panel), $m_X = 20~\mathrm{MeV}$ (top right panel), $m_X = 40~\mathrm{MeV}$ (bottom left panel) and $m_X = 70~\mathrm{MeV}$ (bottom right panel).
The horizontal orange band shows the upper bound on the electron coupling $g_e$ derived from SINDRUM searches for rare charged-pion decays.  
The orange dashed line indicates the expected sensitivity on future searches for the same channel as carried out by the forthcoming PIONEER experiment. The purple dashed line instead represents the expected bound from the future study of Bhabha scattering at the JLab facility\,\cite{Pustyntsev:2025iht}.
The light-blue shaded region shows the current limit on DM-electron scattering from direct detection searches. The horizontal green dashed lines represent constraints from benchmark scenarios for the $g-2$ of the electrons. The lighter (darker) gray regions correspond to limits from beam-dump experiments, namely from the Orsay Collaboration (E137 Collaboration). The red area corresponds to the ensemble of the bounds from direct detection of DM on nucleons, arising from loop-induced contributions, as set by DarkSide-50, XENONnT and LZ experiments. Finally, the vertical dashed black line corresponds to $m_{\psi} = 2m_X$ (see the text for details), while the black area on the left refers to the region $m_{\psi} < m_X$ already studied by Ref.\,\cite{Cesarotti:2024rbh}.
}

 \label{fig:DMparamspace}
 \end{figure*}

Our results are shown in the $(m_\psi,\, g_e)$ plane in Fig.~\ref{fig:DMparamspace} for different values of the mediator masses $m_X$.

Interestingly, for each value of $m_X$, the allowed parameter space in the $(m_\psi,\, g_e)$ plane is restricted to a small, closed region (shown in white in each panel), which shrinks as $m_X$ decreases. This result comes from a non-trivial interplay between bounds from direct detection on electrons and nucleons and constraints from low-energy observables, such as rare pion decays and beam-dump experiments.

For masses $m_X \lesssim m_\psi \lesssim 2m_X$, neither the DM particle nor the mediator can maintain thermal equilibrium with the SM bath up to DM freeze-out, for all values of $g_e$ shown. A proper determination of the relic abundance would therefore require a modified treatment. However, this affects only a very small portion of parameter space. In fact, when $m_\psi < m_X$, the annihilation channel in Eq.~\eqref{eq:XX} is kinematically closed and DM can annihilate only into electrons. As discussed in Section~\ref{sec:model}, this fixes $g_e$ (as a function of $m_\psi$) to values already excluded by direct detection, collider, or BBN constraints.

In conclusion, the complementarity of laboratory and cosmological constraints selects a narrow window for both the DM and mediator masses in the sub-GeV range.
This result is particularly interesting, as next-generation experiments will be able to probe or exclude electrophilic DM models with a scalar mediator of $\mathcal{O}(10~\mathrm{MeV})$ and a sub-GeV Majorana DM particle. A concrete example is provided by the case $m_X = 10$ MeV (top-left panel of Fig.~\ref{fig:DMparamspace}), where the expected improvements from PIONEER in searches for rare pion decays will play a crucial role in testing the remaining viable parameter space.

Finally, as mentioned above, the presence of the electrophilic light mediator motivates exploring a possible connection with the SIDM framework, especially in light of the absence of tree-level hadronic couplings, which arise at two-loops. However, constraints from nuclear recoils are so strong to overcome the two-loop suppression and basically rule out thermally produced DM with mass above $\sim 10$ GeV, as clearly shown in Fig.~\ref{fig:DMparamspace}. Combined with the discussion of the previous sections, this leads to the conclusion that, despite the suppressed coupling to nucleons, a simple electrophilic model is not able to provide a viable SIDM scenario and more complicated model-building strategies should be explored (for instance, along the lines of~\cite{Hambye:2019tjt}).

\subsection{A $X_{17}$ Dark Portal via Freeze-out}
\label{sec:X17OUT}

In this section, we consider the possibility that the putative \(X_{17}\) boson, suggested by hints in the \(e^+e^-\) channel at an invariant mass around 17 MeV reported by the PADME experiment~\cite{PADME:2025dla} and by nuclear transition anomalies observed at ATOMKI, acts as a portal between the SM and a dark sector.\footnote{The current status of the ATOMKI anomalies hints that spin-zero scenarios are exclude as the beryllium transition excludes a CP-even scalar, while the carbon transition disfavors a CP-odd (pseudoscalar) interpretation of \(X_{17}\)~\cite{Feng:2020mbt,Barducci:2022lqd}.
Nevertheless, non-resonant production in \(p + A \to N + e^+ e^-\) can reopen the parameter space for both CP-even and CP-odd scalars, as recently suggested also by Ref.~\cite{Fieg:2026zkg}, thus motivating our analysis.
See Appendix~\ref{app:X17_sub} for a detailed overview of the current experimental situation surrounding the $X_{17}$ puzzle.}
To this purpose, we extend our low-energy interaction Lagrangian in Eq.~\eqref{eq:LAG} to include couplings to quarks as
\begin{equation}\label{eq:LAGq}
    \mathcal{L}_{\rm int}
    = g_{e}\,\bar{e} e \, X
    + \sum_q g_{q}\,\bar{q} q \, X
    + \frac{g_\psi}{2}\,\psi\psi \, X \,,
\end{equation}
where for simplicity we consider only couplings to up and down quarks, namely $q = u,d$, as related to ATOMKI experiment.

While at high energies the $X_{17}$ boson couples directly to quarks as in Eq.~(\ref{eq:LAGq}), at low energies one can instead describe its coupling to nucleons~\footnote{See, for instance, Ref.~\cite{Batell:2018fqo} for a hadrophilic dark matter scenario.}, with interactions of the form $g_N X \bar{N} N$, where $N = \{p,n\}$. In Appendix~\ref{app:chipt}, we briefly review the chiral perturbation theory framework required to derive these effective hadronic couplings, as well as the amplitudes relevant for the scattering processes considered.
Following the analyses of Refs.~\cite{Barducci:2022lqd,Barducci:2025hpg}, nucleon couplings of order $g_N \sim 10^{-3}$ are required for a scalar mediator {with an invariant mass close to \(17\,\mathrm{MeV}\)} to account for the observed nuclear transition anomalies.

We focus on DM masses above 17 MeV, since for lighter masses both production and constraints would be the same as in the electrophilic scenario, already excluded in that mass window.

The inclusion of hadronic couplings does modify significantly both the freeze-out dynamics and the experimental constraints. 
In particular, the presence of such relatively large hadronic couplings completely spoils the results of the previous sections when applied to $X_{17}$, as they induce very large DM-nucleon elastic cross sections. Given the lightness of the mediator and the relatively large couplings, (for both spin-0 and spin-1 mediators), the cross section turns out to be so large that DM masses up to roughly the Planck scale are excluded~\cite{Cirelli:2024ssz}, thus closing all the available parameter space.
Notice that, in some region of the parameter space, the DM mass is below the sensitivity of underground detectors searching for nuclear recoils or the  cross sections are so large that the DM particles never reach underground detectors, so that traditional laboratory constraints cannot be applied. Such large cross sections are however excluded by a combination of different bounds arising from the CMB, structure formation, boosted DM from scatterings on cosmic rays, dynamics of the Milky way satellites and gas heating~\cite{Mahdawi:2018euy,Bringmann:2018cvk,Bramante:2018tos,Bhoonah:2018wmw,Xu:2018efh,Nadler:2019zrb,Maamari:2020aqz,Straight:2026wts}. Thus,
$X_{17}$ boson cannot act as the mediator for fermionic DM produced thermally.

We can relax our assumptions in two possible ways. First, DM could be a scalar particle. 
While most of the constraints also apply in this case, it turns out that BBN bounds rules out thermal scalar DM only for masses above $0.9$ MeV. Thus, in a very  narrow DM mass window, between roughly $1$ and $4$ MeV, scalar DM is not excluded by neither BBN nor DD searches. Notice that such low masses are excluded by CMB constraints unless annihilations are $p$-wave suppressed. For scalar DM the only option to provide a $p$-wave suppressed annihilation is to select a spin-1 $X_{17}$, while a scalar mediator is not viable. Finally, the DM–nucleon elastic scattering cross section is sufficiently suppressed to evade all current constraints. This possibility was already discussed in~\cite{deNiverville:2011it}. While technically viable, it appears rather fine-tuned.

The alternative option is that DM is produced non-thermally. One clean possibility is freeze-in production\,\cite{Hall:2009bx}, which occurs if the coupling of DM to $X_{17}$ is so small that DM particles never reached thermal equilibrium along the cosmological evolution. In the following section we discuss freeze-in production of DM, motivated by the $X_{17}$ anomaly. For completeness, we first discuss the purely electrophilic scenario to underline the differences with the case of thermal production.

\section{Freeze-in}
\label{sec:BoltzmannIN}

Contrarily to the case of freeze-out, we now consider very small values of the coupling $g_\psi$ in Eq.\,\eqref{eq:LAG}, so that DM never thermalized with the SM thermal bath\footnote{We assume that the lepton coupling $g_\ell$ is large enough that the mediator $X$ is kept in thermal equilibrium.}. Assuming a vanishing initial population, DM is produced through decays or annihilations of particles in the thermal bath (SM particles or $X$), which occasionally produce pairs of DM particles. Since the DM density is always very small compared to its thermal equilibrium value, one can solve the relevant Boltzmann equation neglecting the back-reaction term $\propto n_{\psi}^2$. For instance, in the electrophilic scenario with $m_\psi>m_X/2$, the DM particles are produced through $e^+e^-\to X\to \psi\psi$ annihilations and the corresponding number density satisfies the Boltzmann equation

\begin{equation} 
\label{eqBoltzFI}
\dot{n}_\psi+3Hn_\psi\simeq\langle{\sigma v}\rangle_{e^+e^-}n_{e,\rm eq}^2\,,
 \end{equation}
where $n_{e,\rm eq}$ is the equilibrium number density of electrons/positrons.

\subsection{Electrophilic Mediator}
\label{sec:electroFI}

In the electrophilic scenario of Eq.\,(\ref{eq:LAG}), the relevant Boltzmann equation can be solved by treating separately two production channels : $i)$ $X \to \psi\psi$ and $ii)$ $e^+ e^- \to \psi\psi$.
The amplitudes for the channels $i)$ and $ii)$ can be found in Appendix\,\ref{app:ampl}.

For what concerns the channel $i)$, the solution of the Boltzmann equation reads
\begin{equation}
    \frac{\Omega_{DM}h^2}{0.12}  \propto g_\psi^2m_\psi\,\frac{(m_X^2-4m_\psi^2)^{3/2}}{\,m_X^4}\,.
\label{eq:YXcc}
\end{equation}
It is worth empashizing that Eq.\,(\ref{eq:YXcc}) depends only on the coupling $g_\psi$, while being independent of $g_e$. For what concerns, instead, the channel $ii)$, the solution of the Boltzmann equation is
\begin{equation}
\frac{\Omega_{DM}h^2}{0.12} \propto g_\psi^{2}g_e^{2} m_\psi
\int_{s_{\min}}^{\infty}\! ds\;
\frac{\big(s-4m_\psi^{2}\big)^{3/2}\,\big(s-4m_e^{2}\big)^{3/2}}
{s^{5/2}\,\Big[\Gamma_X^{2} m_X^{2}+\big(m_X^{2}-s\big)^{2}\Big]}\,    \,,
\label{eq:Yeecc}
\end{equation}
where $s \equiv (p_{e^-} + p_{e^+})^2$ and $\hat g_{e^{-}},\,\hat g_{e^{+}}$ are initial-state multiplicities of electrons and positrons, respectively. We indicate as $s_{\rm {min}}$ the lowest value that the Mandelstam variable $s$ can have. Note that, contrarily to Eq.\,(\ref{eq:YXcc}), Eq.\,(\ref{eq:Yeecc}) depends on both the couplings $g_\psi$ and $g_e$.

Our phenomenological study can be divided into two distinct regimes in the context of freeze-in, namely:
\begin{itemize}
\item a \emph{non-resonant} regime, in which $m_\psi > m_X/2$ and, thus, only the $e^+\,e^-\to \psi \,\psi$ annihilation is kinematically open. In this case, the lower limit on the integral in Eq.\,(\ref{eq:Yeecc}) is $s_{\rm {min}} = 4m_\psi^2$ and the electrophilic mediator never goes on-shell;
\item a \emph{resonant} regime, where $m_\psi \leq m_X/2$ and the decay $X \to \psi \, \psi$ is kinematically allowed. One can explicitly show that in this case the scattering $e^+\,e^-\to \psi \,\psi$  gives exactly the same contribution of the decay $X \to \psi \, \psi$ to the DM abundance by directly exploiting the Narrow Width Approximation (NWA)\footnote{In general, the NWA can be used only if the decay width of the mediator $X$ is much smaller than its mass, i.e. $\Gamma_X / m_X \;\ll\;1$. To verify that this condition holds, we have assumed that, among all the decay channels of $X$ allowed by the interaction Lagrangian in Eq.\,(\ref{eq:LAG}), $X \to e^+\,e^-$ gives the dominant contribution, thus obtaining that $\Gamma_X / m_X\;\approx\mathcal{O}\!\left(g_e^2/(8\pi)\right)\;\ll\;1 $. This assumption is well justified since the coupling $g_{\psi}$ is expected to be tiny when DM is produced via freeze-in.} inside Eq.\,(\ref{eq:Yeecc})~\cite{Heeck:2014zfa}. The physical reason of this coincidence is that the resonant contribution to the scattering is dominated by the production of an on-shell $X$ from the $e^+e^-$ pair, followed by its decay into DM.
Since in our set-up the electrophilic mediator is assumed to remain in equilibrium with the SM bath, its number density does not depend on the details of how efficiently $e^+e^-$ can resonantly produce $X$ and, a consequence, the only relevant channel for the computation of the DM abundance is the decay $X \to \psi\psi$. 
\end{itemize}
In the case of frozen-in DM production, our findings are that
$ g_\psi\,g_e \sim 10^{-13}$ in the non-resonant regime, while in the resonant one $g_\psi \sim 10^{-13}$. These conclusions are in agreement with the results of the computations carried out in Ref.~\cite{Hall:2009bx}, as well as with the discussion in Section 2 of Ref.\,\cite{Belfatto:2021ats}.

\begin{figure*}[t]
  \centering
  \begin{tabular}{cc}
    \includegraphics[width=0.6\linewidth]{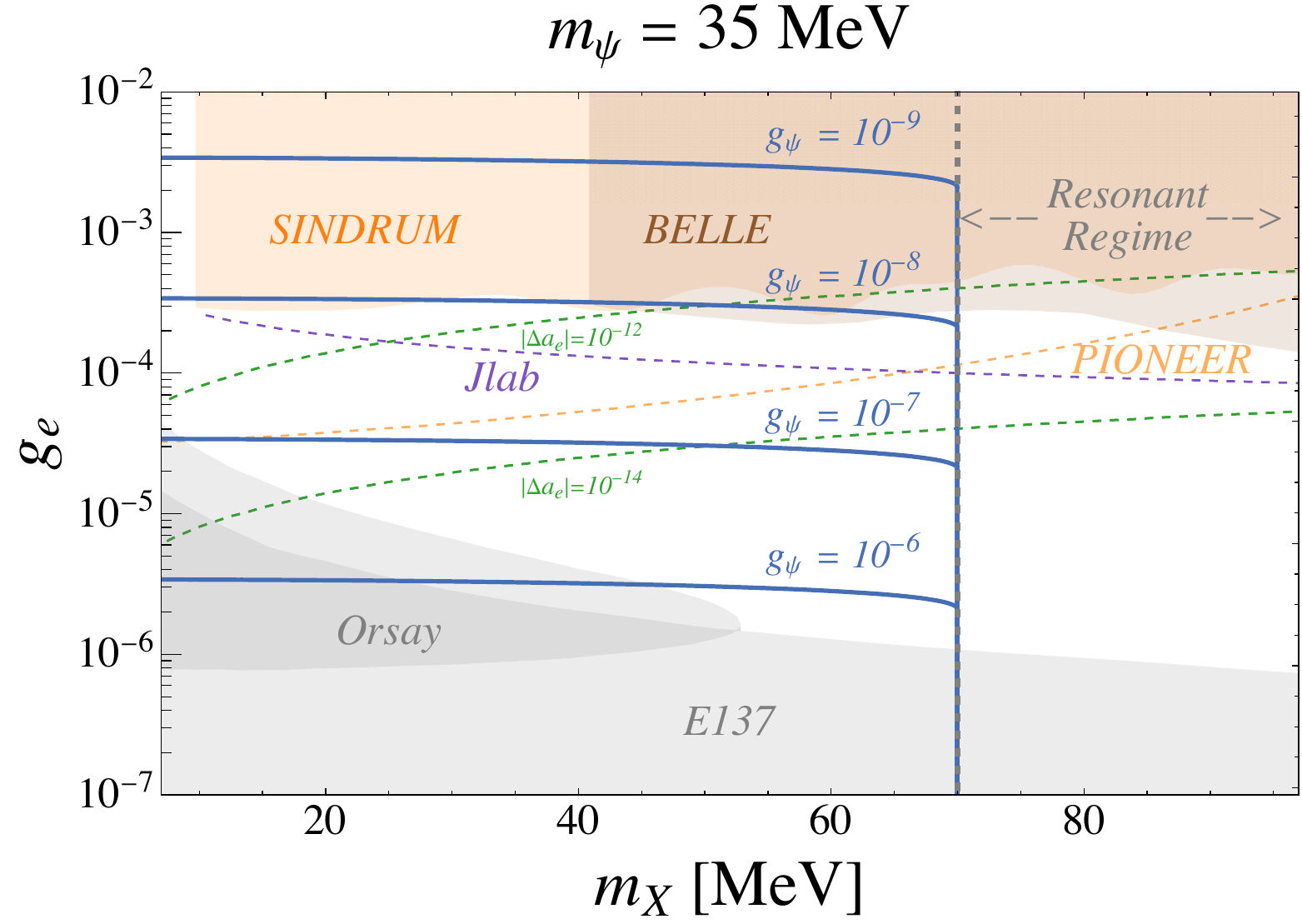}  
  \end{tabular}
  \caption{
      The solid blue lines are the results of the matching of our DM abundance computations in the non-resonant regime onto the observed value of $\Omega_{\rm DM}h^2$ for $m_\psi = 35~\text{MeV}$ and for some benchmark values of the coupling $g_\psi$. Such curves are compared with the exclusion regions and curves recently computed in Ref.\,\cite{DiLuzio:2025ojt} by analyzing low-energy observables, namely: measurements of rare pion decays by SINDRUM experiment (orange solid band) and the future projections by PIONEER (orange dashed line), as well as benchmark scenarios for the $g-2$ of the electrons (green dashed lines) and projected contraints from the future study of Bhabha scattering at JLab (purple dashed line). The gray-shaded regions refer to exclusions from electron beam-dump  experiments, namely Orsay~\cite{Davier:1989wz} and E137~\cite{Bjorken:1988as}, recast in the $(m_X,g_e)$ plane for an electrophilic scalar.
      }
      
  \label{fig:NR_FI}
\end{figure*}

Fig.\,\ref{fig:NR_FI} shows our results on the plane $(m_X, g_e)$ for a benchmark DM mass value $m_{\psi} = 35$ MeV. In the non-resonant regime, the resolution of the Boltzmann equations gives the blue curves, corresponding to the benchmark values $g_\psi = \{10^{-9},\,10^{-8},\,10^{-7}, 10^{-6}\}$. Several of the phenomenological constraints discussed in the previous Section are also illustrated for comparison. The most important message of Fig.\,\ref{fig:NR_FI} is that, quite interestingly, the combination of our relic abundance computation and of the available low-energy experimental constraints yields a quite narrow interval of acceptable values for the coupling $g_\psi$, namely $g_\psi \in [\sim 10^{-8},\,\sim10^{-6}]$. Such an interval may become even narrower in the future, as the projections by next-generation experiments clearly suggest. Such a finding makes our model quite predictive also in the freeze-in scenario, given the fact that our conclusions are not very sensitive to the chosen DM mass. 
{Indeed, changing the value of $m_\psi$ simply shifts the boundary between the resonant and non-resonant regimes, increasing or reducing the range of allowed values of the mediator mass $m_X$, but it does not alter the qualitative behavior of the freeze-in curves.}

The situation is quite different in the resonant regime, where the relic abundance is entirely independent of the coupling $g_e$. A detailed discussion of the relevant findings and numerical details characterizing this scenario can be found in Appendix\,\ref{app:appc}.

\subsection{A $X_{17}$ Dark Portal by Freeze-in}
\label{sec:X17FI}
As already mentioned in Sec.\,\ref{sec:X17OUT}, the possibility of identifying our mediator with the $X_{17}$-boson is excluded in the case of freeze-out scenario due to the impact of the introduction of coupling to nucleons on direct detection. This observation naturally motivates us to scrutinize in what follows whether this scenario (namely the one represented by Eq.\,(\ref{eq:LAGq})) is viable through freeze-in. In line with what argued in Ref.\,\cite{Barducci:2022lqd}, we are going then to consider the case $g_q \gg g_e$, so that the channels involving quarks or hadrons will be the dominant ones in the computation of the relic abundance.

The relevant production processes are  $i)\,\bar qq\to\psi\psi$ and $ii)\,hh\to\psi\psi$ ($h$ denotes a generic hadron, $e.g.$ a pion or a nucleon) at temperatures above and below the QCD crossover, respectively.

For $m_\psi\gg\Lambda_{\rm QCD}$, the dark matter production is dominated by the channel $i)$, namely quark annihilations. This process is most efficient at temperatures  well above the QCD phase transition, where the thermal bath is appropriately described in terms of partonic degrees of freedom. In contrast, hadronic processes only become relevant below the QCD scale, $T\lesssim \Lambda_{\rm QCD}$, and their corresponding annihilation rates are exponentially suppressed as $e^{-2m_\psi/T}$. In the opposite regime, $m_\psi\ll\Lambda_{\rm QCD}$, both quark- and hadron-level processes may, in principle, contribute. However, since the coupling of the $X_{17}$ mediator to hadrons is larger than its coupling to quarks, channel $ii)$ becomes the dominant production mechanism in this case.

\begin{figure*}[t]
  \centering
  \begin{tabular}{cc}
    \includegraphics[width=0.6\linewidth]{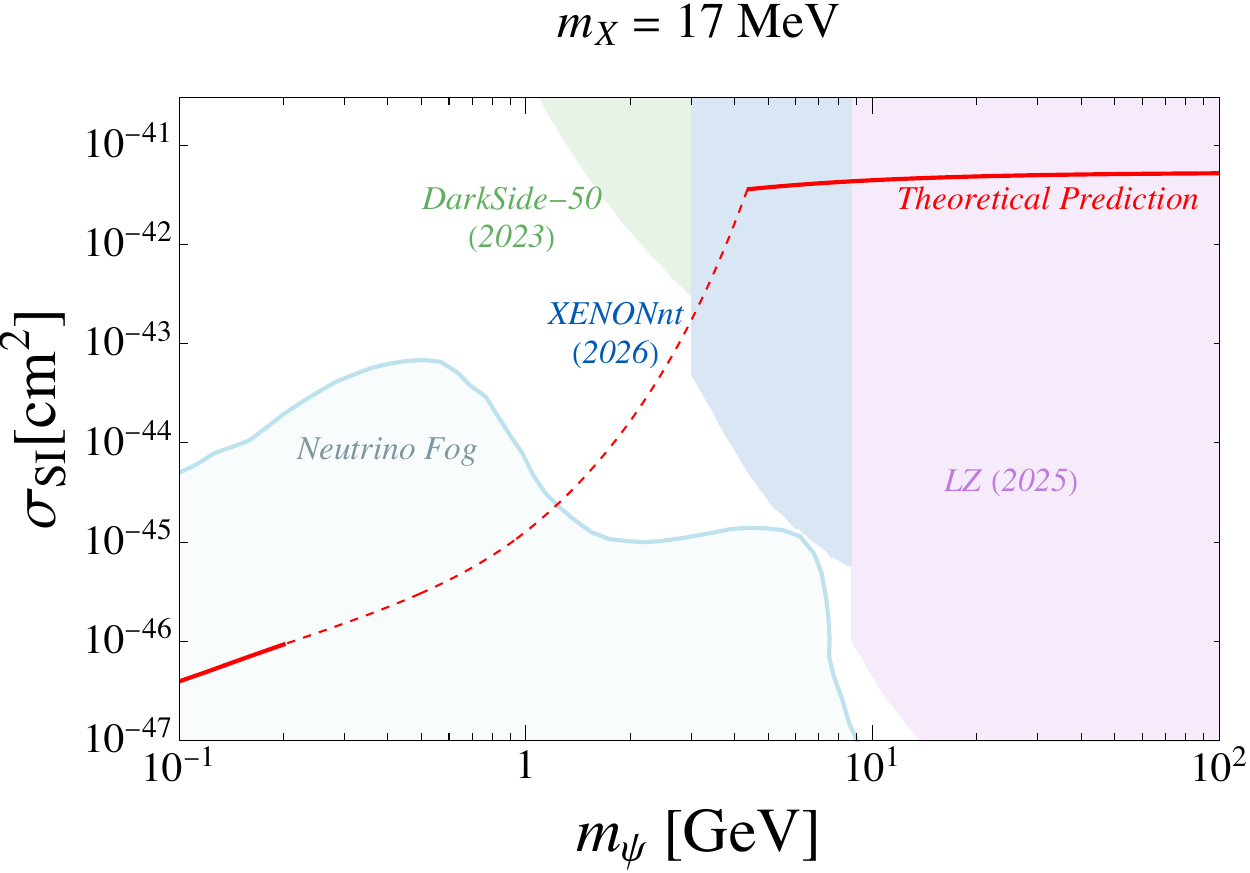}  
  \end{tabular}
  \caption{
      The solid red line is the result of our computation of the SI DM-nucleon elastic cross section as in Eq.\,(\ref{eq:SIFI}), after having matched our DM abundance computations via freeze-in onto the observed value of $\Omega_{\rm{DM}}h^2$. Our result is compared with the exclusion lines corresponding to current constraints from direct detection with nucleons\,\cite{DarkSide:2022knj,XENON:2026qow,LZ:2024zvo} and to the ``neutrino fog'' background\,\cite{OHare:2021utq}. Finally, the red dashed curve represents an approximated interpolation between the low- and the high-mass regimes.
      }
  \label{fig:X17_FI}
\end{figure*}

It is worth noting that there is an intermediate region, approximately between $200\,\mathrm{MeV}$ and few $\,\mathrm{GeV}$, in which both the hadronic and partonic contributions should be in principle included. 
This intermediate region is indicated by the red dashed curve in Fig.~\ref{fig:X17_FI}, which marks the range where our theoretical estimate is not reliable. A more precise computation of all the relevant production channels in this regime would require a dedicated matching between the high-energy and low-energy regimes, which lies beyond the scope of the present work.

What we discuss in the following is, thus, our theoretical predictions for the spin-independent (SI) DM-nucleon elastic cross section $(\sigma_{SI})$ for $m_\psi \ll \Lambda_{\rm QCD}$ and $m_\psi \gg \Lambda_{\rm QCD}$. The expression of $(\sigma_{SI})$ reads~\cite{Batell:2018fqo}
\begin{equation}
\label{eq:SIFI}
    \sigma_{\rm SI}
    =
    \frac{\mu_{\psi N}^{2}}{\pi}\,
    \frac{\left[ Z f_p + (A-Z) f_n \right]^2}{A^2}\,,
\end{equation}
with
\begin{equation}
    f_{p(n)}
    \simeq
    \frac{g_\psi g_q \, m_{p(n)}}{m_X^2}
    \left(
         \frac{F_d^{p(n)}}{m_d}
        +
         \frac{F_u^{p(n)}}{m_u}
    \right)\,.
    \label{eq:SIDMN}
\end{equation}
Here $\mu_{\psi N}$ denotes the nucleon-DM reduced mass,
$A$ and $Z$ are the mass and atomic numbers of the target nucleus, respectively, and $m_u$ and $m_d$ are the $u$ and $d$ quark masses.
The quantities $F_{u,d}^{p(n)}$ are nucleon form factors, for which we adopted the numerical values $F_{u}^{p}=0.023$, $F_{d}^{p}=0.032$, $F_{u}^{n}=0.017$ and $F_{d}^{n}=0.041$\,\cite{Hisano:2010ct}\footnote{As discussed in Appendix B of Ref.\,\cite{Bloch:2024suj}, at present some numerical discrepancies exist among different non-perturbative computations of the nucleonic quantities $F_u^{p(n)},\,F_d^{p(n)}$. We have verified numerically that changing the values of $F_u^{p(n)},\,F_d^{p(n)}$ accordingly does not modify our conclusions.}. 

We stress that the cross section $\sigma_{\rm SI}$ in Eq.\,(\ref{eq:SIFI}) is effectively controlled by the overall factor $g_\psi g_q$. This combination can also be fixed by the requirement of reproducing the observed relic abundance of DM via freeze-in and through all the expressions contained in Appendix \ref{app:chipt}. In particular, in analogy with the findings of Sec.\,\ref{sec:electroFI}, this means that $g_\psi\,g_q \sim 10^{-13}$ for values of DM masses above the QCD scale. In contrast, in the regime $m_\psi\ll\Lambda_{\rm QCD}$, the same combination of couplings is reduced by approximately one order of magnitude, due to the enhanced coupling to hadrons. The results of our theoretical computations are shown in Fig.~\ref{fig:X17_FI} together with the current constraints from direct detection with nucleons\,\cite{DarkSide:2022knj,XENON:2026qow,LZ:2024zvo} and  the ``neutrino fog'' background taken from Ref.\,\cite{OHare:2021utq}.\footnote{Fig.~\ref{fig:X17_FI} only applies to the non-resonant regime. Whenever $m_\psi<m_X$, one recovers the resonant regime discussed in Appendix~\ref{app:appc}.} As can be seen from this plot, despite the smallness of the couplings, $g_qg_\psi\lesssim 10^{-13}$, DD constraints are so strong to rule out DM candidates with mass above a few GeV, even in the freeze-in scenario. Note that the value of  $\sigma_{\rm SI}$ decreases for $m_\psi\ll\Lambda_{\rm QCD}$  due to a combination of two effects: the reduced effective coupling, $g_q g_\psi$, arising from the enhanced coupling to hadrons, and the scaling of the reduced mass $\mu_{\psi N}\sim m_\psi$ for DM masses below that of the nucleon. 
The most interesting window corresponds to DM masses of the order of 1 GeV, where the predicted DM-nucleon cross-section lies between the excluded and the neutrino fog band, indicating that future searches at low-mass regions could probe this freeze-in scenario.
We remind that our computation is not fully reliable in this region and the dashed line in Fig.~\ref{fig:X17_FI} must be understood as an approximate interpolation between the high- and low-mass regimes, meaning that a proper analysis is needed. 
Taken at face values, our results suggest that the putative $X_{17}$ boson can play the role of dark sector mediator for feebly coupled sub-GeV DM. 

\section{Conclusions}
\label{sec:conclusions}

In this work, we have investigated a simple and well‑motivated scenario in which a Majorana dark matter fermion interacts with the Standard Model through a light CP‑even scalar mediator coupled preferentially to electrons. We have performed a comprehensive study of both thermal (freeze‑out) and non‑thermal (freeze‑in) production mechanisms, identifying the regions of parameter space consistent with the observed dark matter relic abundance while satisfying current astrophysical, cosmological, laboratory, and direct‑detection constraints.

Our analysis shows that dark matter interacting through a light electrophilic mediator can reproduce the correct relic abundance via freeze‑out only in a limited mass range, roughly from $\mathcal{O}(10)\,\mathrm{MeV}$ up to a few GeV. Even within this window, however, the viable parameter space is strongly constrained by direct‑detection experiments sensitive to electron recoils, as well as by accelerator searches for light mediators. As a consequence, thermally produced electrophilic dark matter is increasingly challenged by existing data.

We have also examined the possibility of realizing self‑interacting dark matter within the same purely electrophilic framework. We find that, once all experimental, astrophysical, and cosmological bounds are consistently taken into account, electrophilic interactions alone cannot generate sufficiently large self‑interaction cross sections without violating existing constraints. This disfavors purely electrophilic models as viable explanations of the small‑scale structure anomalies.

Motivated by recent low‑energy experimental hints, we further explored the scenario in which the scalar mediator is identified with the putative $X_{17}$ state, potentially linking the excess observed by PADME in the $e^+e^-$ channel with the nuclear transition anomalies reported by the ATOMKI collaboration. In this framework, the inclusion of hadronic couplings is required to account for the nuclear data. We have shown that, once such couplings are introduced, dark matter production through freeze‑out is generically ruled out, while viable scenarios persist only through the freeze‑in mechanism. These solutions are restricted to sub‑GeV dark matter masses and feeble mediator couplings. Notably, while previous studies have mostly focused on vector‑mediator realizations of the $X_{17}$ hypothesis, our work demonstrates that a light CP‑even scalar provides a qualitatively distinct and viable alternative  as the dark sector mediator of feebly coupled sub-GeV DM. 

Taken together, our results point toward light, sub‑GeV dark matter interacting through very weak portals to the Standard Model as a particularly
well‑motivated and experimentally accessible class of scenarios. They also highlight the crucial role of experimental complementarity: direct‑detection searches targeting electron recoils and laboratory experiments probing light mediators provide synergistic and often decisive constraints on this class of models.
In this sense, our work can be viewed as the light‑mediator counterpart of Ref.~\cite{Cesarotti:2024rbh}, extending its conceptual framework to the regime where the mediator is light rather than heavy. Future progress in precision low‑energy experiments and dedicated dark matter searches will be essential to further probe this remaining parameter space and clarify the nature of possible light‑mediator signals.

\section{Acknowledgements}
We would like to thank Fabio Bossi and Mauro Raggi for  information on the $g_e$ sensitivities, as well as Luca di Luzio and Lorenzo Calibbi for a careful reading of the manuscript and their useful comments. We also thank Roberto Contino and Luca Silvestrini for useful discussions on our DM model and on low-energy observables. The work of FM  is supported by the European Union-Next Generation EU and by the Italian Ministry of University and Research (MUR) via the
PRIN 2022 project No. 2022K4B58X-AxionOrigins.
The work of GL is supported by the INFN Cabibbo Fellowship, call 2025.  This article is based upon work from COST Action COSMIC WISPers CA21106, supported by COST (European
Cooperation in Science and Technology).
The work of CT has received funding from the French ANR, under contracts ANR-19-CE31-0016 (`GammaRare') and ANR-23-CE31-0018 (`InvISYble'), that he gratefully acknowledges.
The work of LV is supported by the Italian Ministry of University and Research (MUR)
and by the European Union's NextGenerationEU program under the Young Researchers 2024
SoE Action, research project \textit{SHYNE}, ID:\,SOE$\_$20240000025.
CT and MG thank the ECT$^*$ institute for hospitality during the final stage of this work.

\appendix

\section{Possible UV completion of our model}\label{app:UV}

In this Appendix we present a simple UV completion of the Lagrangian in 
Eq.~\eqref{eq:LAG}, closely related to the model discussed in Appendix~B of 
Ref.\,\cite{Ema:2020fit}. At the renormalizable level, the dark-sector 
interactions are 
\begin{equation}
    \mathcal{L}_{\psi}
    = \frac{1}{2}\psi^\dagger\, i \sigma^\mu \partial_\mu \psi 
    - \frac{m_\psi}{2}\, \psi \psi
    + \frac{g_\psi}{2}\, X\, \psi \psi \, .
\end{equation}

The interaction with electrons appearing in \eqref{eq:LAG}, 
$g_{eX}\, X_{17}\, \bar e e$, is not invariant under the electroweak gauge 
symmetry. It should therefore be understood as arising from a higher-dimensional 
operator in the theory above the electroweak scale. A minimal choice is
\begin{equation}
\label{eq:dim5_Xee}
\mathcal{L}_{EW} \supset 
-\frac{g_{Xee}}{\Lambda}\,
X\, \bar L H e_R
+ \text{h.c.}\,,
\end{equation}
which, after electroweak symmetry breaking, reproduces the low-energy interaction 
in Eq.~\eqref{eq:LAG}.
The presence of the non-renormalizable operator in Eq.~\eqref{eq:dim5_Xee} 
implies, as usual, the existence of additional heavy degrees of freedom in the UV 
theory. As a minimal completion, we introduce a vector-like pair of fermions with 
the same SM quantum numbers as the right-handed electron,
$E_{L,R}\sim (1,1)_{-1}$.
The renormalizable Lagrangian reads
\begin{equation}
\begin{split}
     \mathcal{L}_E \supset\;& 
     \bar{L} Y_e H e_R 
     + \bar{L} y_E H E_R
     + \mu_E\, \bar{E}_L e_R 
     + M_E\, \bar{E}_L E_R \\
     & + \lambda_E\, X\, \bar{E}_L e_R
     + \lambda'_E\, X\, \bar{E}_L E_R
     + \text{h.c.}\, .
\end{split}
\end{equation}
In the limit $M_E \gg y_E v_H \gg \mu_E$, the fields $(E_L, E_R)$ form a heavy 
vector-like fermion with mass 
$M_E + \mathcal{O}(y_E^2 v_H^2 / M_E)$. 
Integrating out this state at tree level generates the effective electron coupling 
\[
g_{eX} \sim \frac{\lambda_E\, y_E\, v_H}{\sqrt{2}\, M_E}\, .
\]
For a representative choice of parameters 
$M_E \gtrsim 100~\text{TeV}$ and $\lambda_E\,y_E\lesssim0.01$, the desired value of $g_{eX} \sim 10^{-5}$ is produced. 
Moreover, for $M_E \gtrsim 100$~TeV the new states evade LHC searches, and the induced modification of the Higgs–electron coupling remains safely below the current experimental bounds reported by ATLAS\footnote{
Parameterizing the Higgs coupling to electrons as 
$\kappa_e m_e\, h\, \bar{e} e$, one finds 
$\kappa_e \simeq 1 + y_E^2 v_H^2/(m_e M_E)$, which satisfies the present limit 
$\kappa_e \lesssim \mathcal{O}(300)$ from ATLAS. 
For future sensitivities see Refs.~\cite{Erdelyi:2025axy, Allwicher:2025mmc}.}.

We have thus provided a minimal ultraviolet completion of the electron coupling, 
which successfully reproduces the effective interaction in 
Eq.~\eqref{eq:LAG} while remaining compatible with current bounds on new physics.

\section{Physical amplitudes}
\label{app:ampl}

In this Appendix we collect the expressions of the square amplitudes of interest for the computations developed in Sections \ref{sec:BoltzmannOUT} and \ref{sec:BoltzmannIN}. 
For what concerns DM decays, we have simply that:

\begin{align}
  \sum_{\text{pol.}}\left|\mathcal{A}(X \to \psi\psi)\right|^{2}
  &= 2\,g_{\psi}^{2}\,\bigl(m_{X}^{2}-4m_{\psi}^{2}\bigr),
  \label{eq:MXtocc}
\end{align}

For what concerns, instead, the annihilations, the corresponding expressions read:
\begin{align}
  \sum_{\text{pol.}}\left|\mathcal{A}(e^{+}e^{-}\to \psi\psi)\right|^{2}
  &= 
  \frac{4\,g_{\psi}^{2}g_{e}^{2}\,\bigl(s-4m_{e}^{2}\bigr)\,
        \bigl(s-4m_{\psi}^{2}\bigr)}
       {\bigl(m_{X}^{2}-s\bigr)^{2}+m_{X}^{2}\Gamma_{X}^{2}},
  \label{eq:MeetoccD}
\end{align}
where the finite width of the mediator in the $s$-channel propagator has been properly included in the denominators. 

The amplitude for the process $XX \to \psi\psi$ results from the interference between the $t$-channel and $u$-channel contributions and the expression reads:
\begin{equation}
\label{eq:XXpp}
\begin{split}
    \sum_{\text{pol.}}\left|\mathcal{A}(XX\to \psi\psi)\right|^{2}
    =&\frac{2 g_\psi^{4}}{(m_\psi^{2}-t)^{2}(m_\psi^{2}-u)^{2}}\Bigl[-32 m_\psi^{8} + \\
    &G_{6} m_\psi^{6} + G_{4} m_\psi^{4} + G_{2} m_\psi^{2} + \\ 
    &(t-u)^{2}\bigl(tu - m_X^{4}\bigr)\Bigr] \ ,
\end{split}    
\end{equation}
where we defined
\begin{align}
G_{6} &\equiv 16\left(2 m_{X}^{2} + t + u\right) \ , \nonumber \\
G_{4} &\equiv -32 m_{X}^{2}(t+u) + t^{2} + 30tu + u^{2} \ , \\
G_{2} &\equiv (t+u)\left[8 m_{X}^{2}(t+u) -t^{2} - 14tu - u^{2}\right] \ , \nonumber
\end{align}
with $s$, $t$ and $u$ the usual Mandelstam variables.

Let us highlight that the aforementioned processes are the only ones involving the dark fermion $\psi$ and varying the number of DM particles between the initial and final states at least by one unit. Moreover, all the above square amplitudes are summed over all initial and final state polarizations (and \emph{not} averaged over the initial state polarizations).

\section{Freeze-in in the resonant regime for an electrophilic mediator}
\label{app:appc}

In this Appendix we discuss our findings and some interesting numerical details concerning freeze-in computation in the resonant regime, namely when $m_\psi < m_X/2$. In this case, DM production is dominated by the on-shell decay $X\to\psi\psi$, as discussed in Sec.~\ref{sec:electroFI}, and the relic abundance is thus independent of the electron coupling $g_e$. In other words, it depends only on $m_\psi$, $g_\psi$ and $m_X$.

We recall that in the resonant regime one finds
\begin{equation}
    \frac{\Omega_{DM}h^2}{0.12}\propto \,\frac{m_\psi\,(m_X^2-4m_\psi^2)^{3/2}}{m_X^4}\,.
\end{equation}
If one focuses on the mediator mass $m_X$, this scaling shows the interplay of two competing effects. On the one hand, increasing $m_X$ enhances the decay width $X\to\psi\psi$, and therefore tends to increase the DM abundance. On the other hand, freeze-in production occurs at temperatures of order $T\sim m_X$, so larger values of $m_X$ correspond to earlier production times, when the Hubble expansion rate is larger and the produced DM is more strongly diluted. 

\begin{figure*}[t]
  \centering
  \begin{tabular}{cc}
    \includegraphics[width=0.5\linewidth]{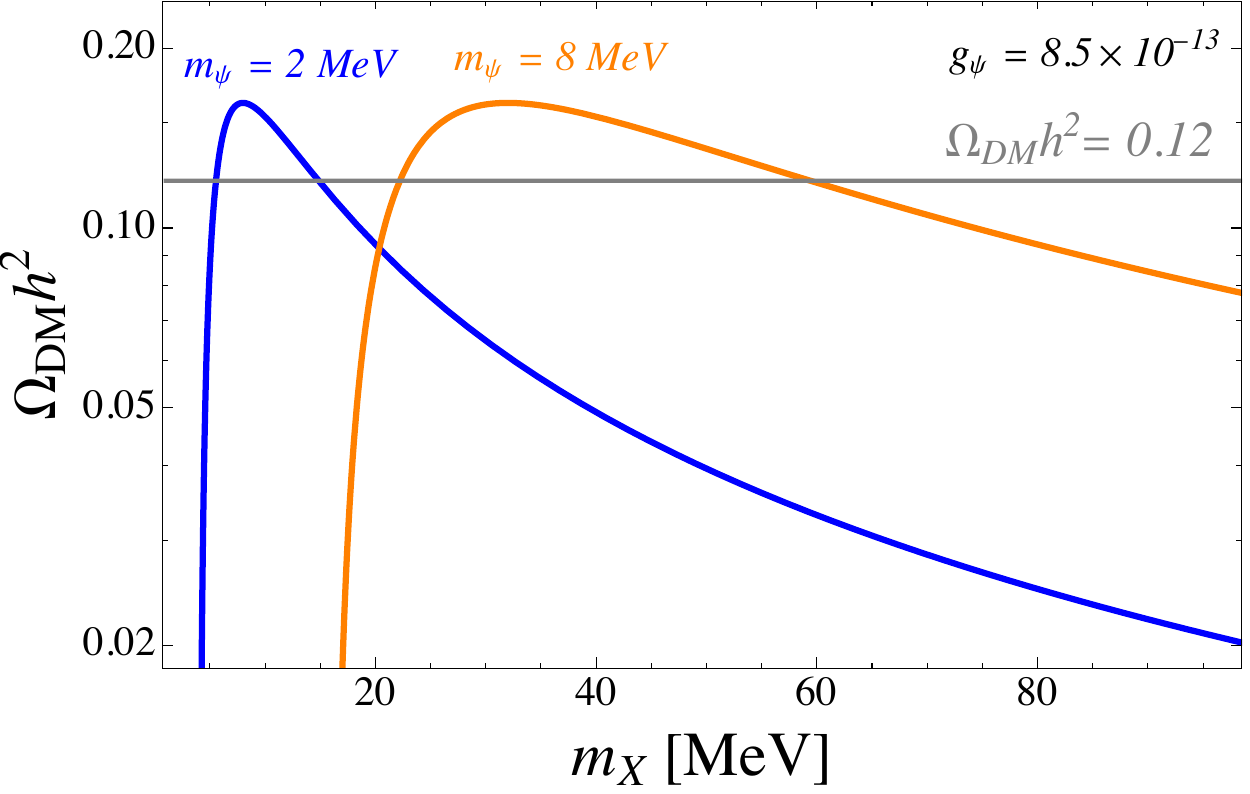} &
    \includegraphics[width=0.5\linewidth]{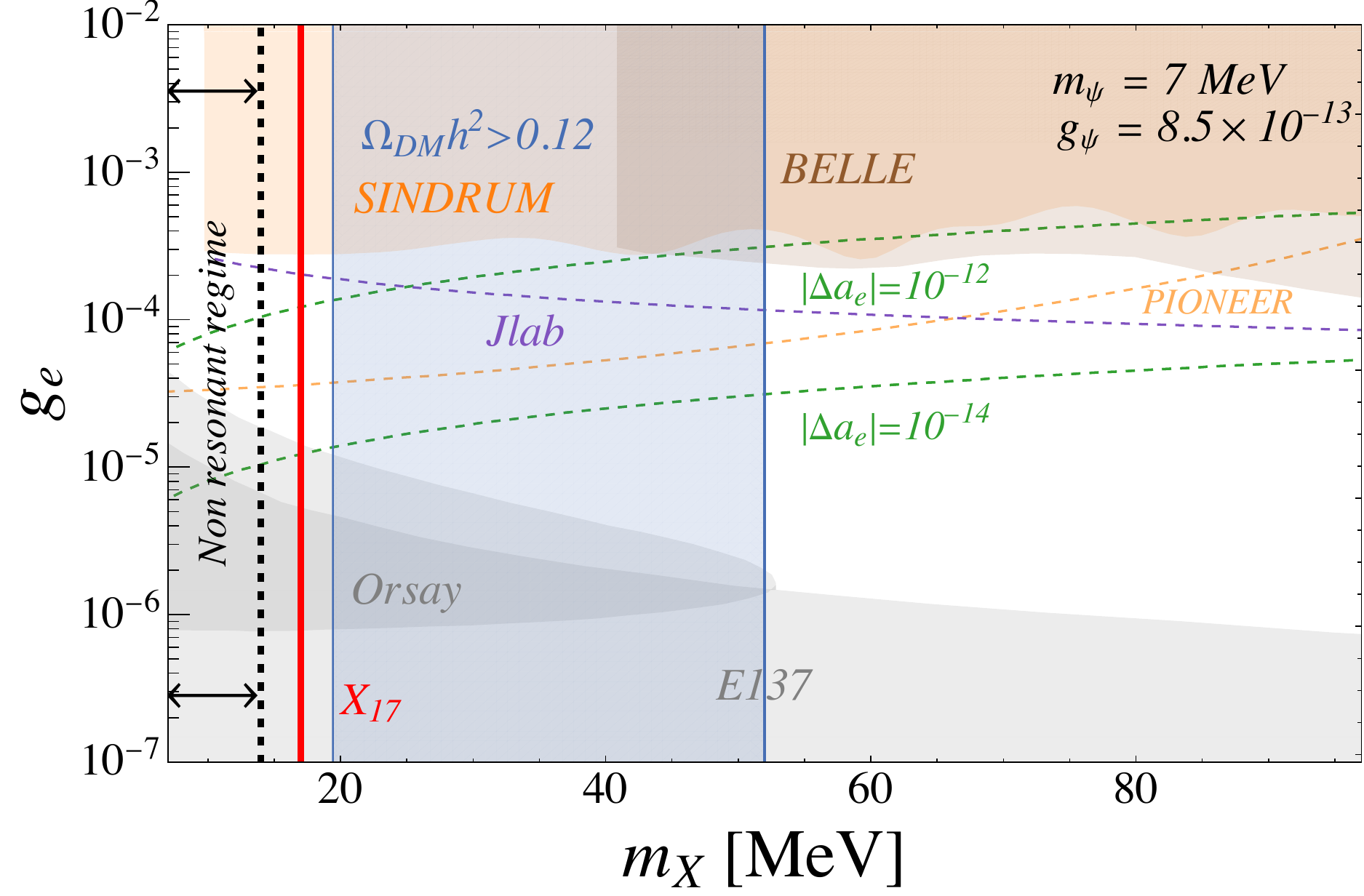} 
  \end{tabular}
  \caption{
      \textbf{\it{Left}}: Relic abundance $\Omega_{\psi}h^2$ in the resonant regime as a function of the mediator mass $m_X$. The coupling $g_\psi$ is fixed to the value $8.5\times 10^{-13}$. The coloured curves correspond to different choices of the DM mass, namely $m_\psi =$\,2\,(blue)\,and\,8\,(orange)\,$\mathrm{MeV}$. The horizontal gray line denotes the observed value of the DM relic abundance, while the vertical red line corresponds to $m_X =17~\mathrm{MeV}$ and is associated with the $X_{17}$ hypothesis.
       {\textbf{\it{Right}}}: Plot analogous to Fig.~\ref{fig:NR_FI} for $m_\psi = 7~MeV$ and $g_{\psi} = 8.5 \times 10^{-13}$ but in the resonant regime, with the vertical blue band corresponding to the requirement of not overproducing DM. See the caption of Figure \ref{fig:NR_FI} for a detailed description of the low-energy constraints.}
  \label{fig:R_RelAb}
\end{figure*}

Because of this competition, the relic abundance is not a monotonic function of $m_X$. This behaviour is shown by the colored curves in the left panel of Fig.~\ref{fig:R_RelAb}, which are obtained by solving the Boltzmann equation for different values of the DM mass $m_\psi$, while fixing the dark-sector coupling to the benchmark value $g_\psi=8.5\times10^{-13}$. The abundance $\Omega h^2$ typically grows just after the kinematic threshold of the resonant regime, $m_X=2m_\psi$, then reaches a maximum, and finally decreases again at larger mediator masses. As a result, the condition $\Omega_\psi h^2 \simeq 0.12$ (represented by the gray horizontal line) can admit two distinct solutions for $m_X$ at fixed value of $m_{\psi}$: the first one is close to threshold, namely $m_X \gtrsim 2m_\psi$, where the abundance is mainly driven by the kinematic growth of the decay rate; the second one corresponds to a larger value of $m_X$, namely in the region where the final abundance is influenced by cosmological dilution.

The features shown in the left panel of Fig.~\ref{fig:R_RelAb} have an important phenomenological consequence. Since in the resonant regime the relic abundance does not depend on $g_e$, fixing the observed DM density does not select a curve in the $(m_X,g_e)$ plane, but rather one or more vertical lines. Furthermore, one may relax the assumption that the $X$-mediated DM production via freeze-in accounts for the whole DM abundance. In other words, if we only require that it does not overproduce DM, a region of the ${m_X,\,g_e}$ plane will be excluded by such vertical bands. For the sake of completeness, in the right panel of Fig.~\ref{fig:R_RelAb}, where the blue area corresponds to the region ruled out by DM overproduction for the benchmark choice $m_\psi=7~\mathrm{MeV}$ and $g_\psi=8.5\times10^{-13}$. In this case, the value $m_X\simeq 17~\mathrm{MeV}$ lies in the narrow branch close to threshold. 

\section{$X_{17}$ status and future direction}
\label{app:X17_sub}
The ATOMKI experiment consists of a proton beam impinging on a target nucleus \(A\) at rest, with the goal of producing an excited nucleus \(N^\star\) and measuring its internal pair creation (IPC) transition to the ground state \(N\), namely
\begin{equation}
p + A \to N^\star \to N + e^+ e^- \, .
\end{equation}
Such transitions have been studied for \(N = {}^{8}\mathrm{Be}\)~\cite{Krasznahorkay:2015iga,Krasznahorkay:2018snd},
\({}^{4}\mathrm{He}\)~\cite{Krasznahorkay:2019lyl,Krasznahorkay:2021joi}, and
\({}^{12}\mathrm{C}\)~\cite{Krasznahorkay:2022pxs}.
For each of these transitions, the ATOMKI collaboration reported an excess of IPC events that is kinematically consistent with the emission of an on-shell boson in the nuclear transition, promptly decaying into an \(e^+ e^-\) pair, with an inferred mass of approximately \(17~\mathrm{MeV}\)~\cite{Feng:2020mbt,Barducci:2022lqd}.

More recently, the MEG~II (Muon Electron Gamma) experiment at the Paul Scherrer Institute~\cite{MEGII:2024urz} investigated the same
\({}^7\mathrm{Li}(p,e^+e^-){}^8\mathrm{Be}\) reaction in order to test the ATOMKI excess.
MEG~II reported no statistically significant evidence for the presence of an \(X_{17}\) particle.
Nevertheless, the results remain consistent with the ATOMKI observations within \(1.5\,\sigma\).
This implies that, while the \({}^8\mathrm{Be}\) anomaly is disfavored, it is not conclusively ruled out, as emphasized in Ref.~\cite{Barducci:2025hpg}.

A central issue concerns the spin and parity quantum numbers of the hypothetical \(X_{17}\) boson.
Under the assumption that \(X_{17}\) is produced exclusively in the decay of the excited nuclear state \(N^\star\), angular momentum and parity conservation impose stringent selection rules on the transition \(N^\star \to N + X_{17}\).
Consequently, the viability of the \(X_{17}\) interpretation depends on whether the observed nuclear transitions are dynamically allowed for a given spin--parity assignment.

For spin-zero scenarios, the ATOMKI data place particularly strong constraints.
The excess observed in the Beryllium transitions excludes a CP-even scalar interpretation of \(X_{17}\), while the Carbon transition rules out a pseudoscalar assignment, as discussed for example in Refs.~\cite{Feng:2020mbt,Barducci:2022lqd}.
Other spin--parity assignments are generally disfavored by tensions among different ATOMKI results and by independent experimental bounds, such as those from NA48~\cite{NA482:2015wmo} and SINDRUM~\cite{SINDRUM:1986klz}, see e.g. Refs.~\cite{Feng:2020mbt,Barducci:2022lqd,Hostert:2023tkg,Mommers:2024qzy,Barducci:2025hpg,Fieg:2026zkg,Hostert:2026xul}.

It is worth noting, however, that the \(X_{17}\) particle might also be produced through non-resonant contributions to the process
\(p + A \to N + e^+ e^-\).
This possibility reopens the parameter space for both CP-even and CP-odd spin-zero particles.
Motivated by this observation, we consider in this work the case of a CP-even, spinless \(X_{17}\) boson.

Concerning its leptonic interactions, the ATOMKI experiment is sensitive to the branching ratio
\(\mathcal{B}(X_{17} \to e^+ e^-)\), rather than to the absolute strength of the electron coupling itself.
If \(X_{17}\) decays exclusively into an \(e^+ e^-\) pair, one has
\(\mathcal{B}(X_{17} \to e^+ e^-)=1\), and the explicit dependence on the electron coupling drops out.
However, the requirement that \(X_{17}\) decays promptly within the detector volume imposes a lower bound on its decay width, which can be expressed as~\cite{Barducci:2025hpg}
\begin{equation}
\label{eq:ULgamma}
\Gamma(X_{17} \to e^+ e^-) \gtrsim 1.3 \times 10^{-10}
\times \mathcal{B}(X_{17} \to e^+ e^-) \, .
\end{equation}
If \(X_{17}\) decays solely into \(e^+ e^-\), this translates into a lower bound on the electron coupling.

An independent and complementary probe of the \(X_{17}\) hypothesis is provided by the PADME experiment, a fixed-target setup in which a positron beam impinges on a thin diamond target.
In this configuration, light new bosons can be searched for via their direct production in purely leptonic processes.
Unlike nuclear experiments, PADME does not rely on nuclear transitions and therefore provides direct sensitivity to the coupling between \(X_{17}\) and electrons.

By varying the energy of the incident positron beam, PADME has explored the center-of-mass energy range
\(14 \lesssim \sqrt{s} \lesssim 23~\mathrm{MeV}\),
which includes the region around \(\sqrt{s} \simeq 17~\mathrm{MeV}\) where \(X_{17}\) production is expected.
Within this window, PADME reported a mild excess in the \(e^+ e^-\) channel with an invariant mass around
\(m_X \simeq 17~\mathrm{MeV}\).
Although the statistical significance of this excess is limited, it is compatible with the mass scale hinted at by the ATOMKI anomalies.

Assuming a vector mediator, PADME interpreted this excess by performing a fit to the data and extracted an upper limit on the electron coupling,
\(g_e \lesssim 5.6 \times 10^{-4}\),
which exceeds the expectation from the background-only hypothesis.
This result represents an additional, purely leptonic indication---kinematically consistent with ATOMKI---for the possible existence of a new state, and allows for a more precise determination of its mass~\cite{Arias-Aragon:2025wdt}.
For a discussion of this result in connection with other electron observables, such as the anomalous magnetic moment, we refer to Ref.~\cite{DiLuzio:2025ojt}.

\section{Hadron interactions for a light mediator}
\label{app:chipt}

In this Appendix we discuss the interactions of the $X_{17}$ boson with hadrons, as arising at low-energy from the quark couplings in the Lagrangian in Eq.\,(\ref{eq:LAGq}). Let us highlight that in the evaluation of our theoretical predictions of $\sigma_{SI}$ in Fig.\,\ref{fig:X17_FI} we included both mesons (pions and kaons) and baryons (nucleons). Given that the interaction among $X$ and the nucleons is the same one described by Eq.\,(\ref{eq:SIDMN}), in what follows we are going to focus only on a $SU(3)$ chiral Lagrangian $\mathcal{L}_{\chi pt}$ accounting for the couplings among $X$ and pions and kaons, respectively.

A CP-even scalar $X$ coupled to $u$ and $d$ quarks with a coupling $g_q$ ($q=u,\,d$) will induce a low energy Lagrangian of the form~\cite{Batell:2018fqo}
\begin{equation}
    \begin{aligned}
        \mathcal{L}_{\chi pt} =&
        \frac{f^{2}}{4}\operatorname{tr}\!\left[D_{\mu}\Sigma\,D^{\mu}\Sigma^{\dagger}\right]\\
        &+ B\frac{f^{2}}{2}\left\{\operatorname{tr}\!\left[\Sigma^{\dagger}\left(m_{q}+X\, c_{q}\right)\right]+\mathrm{h.c.}\right\}
    \end{aligned}
    \label{eq:lagchi}
\end{equation}
$\Sigma(x)=e^{2 i\pi/f}$ contains the meson octet, $f \simeq 93$ MeV is the pion decay constant, $m_q$ is the $3\times 3$ quark mass matrix and $B \simeq 2.6$ GeV is a constant proportional to the chiral condensate. The couplings between the $X$ boson and the light quarks are contained in the matrix
\begin{equation}
    c_q =
\begin{pmatrix}
g_u & 0 & 0 \\
0 & g_d & 0 \\
0 & 0 & 0
\end{pmatrix}.
\end{equation}

From Eq.\,(\ref{eq:lagchi}) we compute the following squared amplitudes 
\begin{align}
\sum_{\text{pol.}}\left|\mathcal{A}(\pi\pi\to \psi\psi)\right|^2
&= \frac{2 (g_d+g_u)^2g_\psi^2 B^2 (s-4m_\psi^2)}{(m_X^2-s)^2} \, , \\
\sum_{\text{pol.}}\left|\mathcal{A}( K_0 \bar K_0\to \psi\psi)\right|^2
&= \frac{2g_d^2 g_\psi^2 B^2 (s-4m_\psi^2)}{(m_X^2-s)^2} \, ,\\
\sum_{\text{pol.}}\left|\mathcal{A}(K^+K^-\to \psi\psi)\right|^2
&= \frac{2g_u^2 g_\psi^2 B^2 (s-4m_\psi^2)}{(m_X^2-s)^2} \, ,
\end{align}
where $\pi\pi\to\psi\psi$ refers to the scattering of both charged and neutral pions.
Such expressions are directly used in Sec.\,\ref{sec:X17FI} for computing the DM relic abundance in the context of freeze-in. In particular, such computations are aimed at fixing the value of the product of couplings $g_{\psi}g_q$ (after having assumed for simplicity $g_u = g_d \equiv g_q$), which is a fundamental ingredient to draw the red lines in Fig.\,\ref{fig:X17_FI}.

\bibliographystyle{apsrev4-1}
\bibliography{bibliography}

\end{document}